\documentclass[manuscript)]{aastex}

\usepackage{lscape,graphicx,epsfig,natbib,color}
\usepackage{epstopdf}                
\makeatletter

\newcommand{\Rmnum}[1]{\expandafter\@slowromancap\romannumeral #1@}
\makeatother

\slugcomment{Submitted to ApJ}

\shorttitle{A partially eruptive filament}
\shortauthors{Q. M. Zhang et al.}

\begin{document}
\title{Multiwavelength observations of a partially eruptive filament on 2011 September 8}

\author{Q. M. Zhang\altaffilmark{1}, Z. J. Ning\altaffilmark{1}, Y. Guo\altaffilmark{2,3},
T. H. Zhou\altaffilmark{1}, X. Cheng\altaffilmark{2,3}, H. S. Ji\altaffilmark{1}, L. Feng\altaffilmark{1,4},
and T. Wiegelmann\altaffilmark{4}}

\affil{$^1$ Key Laboratory for Dark Matter and Space Science,
Purple Mountain Observatory, CAS, Nanjing 210008, China\email{zhangqm@pmo.ac.cn}}
\affil{$^2$ School of Astronomy and Space Science, Nanjing University, Nanjing 210093, China}
\affil{$^3$ Key Laboratory for Modern Astronomy and Astrophysics (Nanjing University),
Ministry of Education, Nanjing 210093, China}
\affil{$^4$ Max-Planck-Institut f\"{u}r Sonnensystemforschung,
Justus-von-Liebig-Weg-3, 37077 G\"{o}ttingen, Germany}

\begin{abstract}
  In this paper, we report our multiwavelength observations of a partial filament eruption event in NOAA
  active region 11283 on 2011 September 8. A magnetic null point and the corresponding spine and
  separatrix surface are found in the active region. Beneath the null point, a sheared
  arcade supports the filament along the highly complex and fragmented polarity inversion line. After
  being activated, the sigmoidal filament erupted and
  split into two parts. The major part rose at the speeds of 90$-$150 km s$^{-1}$ before
  reaching the maximum apparent height of $\sim$115 Mm. Afterwards, it returned to the solar surface
  in a bumpy way at the speeds of 20$-$80 km s$^{-1}$. The rising and falling motions were clearly
  observed in the extreme-ultravoilet (EUV), UV, and H$\alpha$ wavelengths. The failed eruption
  of the main part was associated with an M6.7 flare with a single hard X-ray source. The 
  runaway part of the filament, however, separated from and rotated around the major part for $\sim$1 
  turn at the eastern leg before escaping from the corona,
  probably along large-scale open magnetic field lines. The ejection of the runaway part resulted in
  a very faint coronal mass ejection (CME) that propagated at an apparent speed of 214 km s$^{-1}$  
  in the outer corona.
  The filament eruption also triggered transverse kink-mode oscillation of the adjacent coronal loops in
  the same AR. The amplitude and period of the oscillation were 1.6 Mm and 225 s. 
  Our results are important for understanding the mechanisms of partial filament eruptions 
  and provide new constraints to theoretical models. The multiwavelength observations also shed 
  light on space weather prediction.
\end{abstract}

Online-only material: animations, color figures

\section{Introduction}
Solar prominences or filaments are cool and dense plasmas embedded in the million-Kelvin corona \citep{mac10}.
The plasmas originate from the direct injection of chromospheric materials into a preexisting filament channel,
levitation of chromospheric mass into the corona, or condensation of hot plasmas from the chromospheric evaporation
due to the thermal instability \citep{xia11,xia12,rony14,zhou14}. Prominences are generally believed to be supported
by the magnetic tension force of the dips in sheared arcades \citep{guo10b,ter15} or twisted magnetic flux ropes
\citep[MFRs;][]{su12,sun12a,zhang12a,cx12,cx14a,xia14a,xia14b}. They can keep stable for
several weeks or even months, but may get unstable after being disturbed. Large-amplitude and long-term filament
oscillations before eruption have been observed by spaceborne telescopes \citep{chen08,li12,zhang12b,bi14,shen14}
and reproduced by numerical simulations \citep{zhang13}, which makes filament oscillation another precursor for
coronal mass ejections \citep[CMEs;][]{chen11} and the accompanying flares. When the twist of a flux rope supporting
a filament exceeds the threshold value (2.5$\pi$$-$3.5$\pi$), it will also become unstable and erupt due to the ideal kink
instability \citep[KI;][]{hood81,kli04,tor04,tor10,fan05,sri10,asch11,kur12}.
However, whether the eruption of the kink-unstable flux rope
becomes failed or ejective depends on how fast the overlying magnetic field declines with height \citep{tor05,liu08a,kur10}.
When the decay rate of the background field exceeds a critical value, the flux rope will lose equilibrium and erupt via
the so-called torus instability \citep[TI;][]{kli06,jiang14,ama14}. On the other hand, if the confinement from the
background field is strong enough, the filament will decelerate to reach the maximum height before falling back to the
solar surface, which means the eruption is failed \citep{ji03,liu09,guo10a,kur11,song14,jos13,jos14}.

In addition to the successful and failed eruptions, there are partial filament eruptions \citep{gil07,liu07}. After examining 54
H$\alpha$ prominence activities, \citet{gil00} found that a majority of the eruptive prominences show separation of
escaping material from the bulk of the prominence; the latter initially lifted away from and then fell back to the solar
surface. To explain the partial filament eruptions, the authors proposed a cartoon model in which magnetic reconnection
occurs inside an inverse-polarity flux rope, leading to the separation of the escaping portion of the prominence and the
formation of a second X-type neutral line in the upper portion of the prominence.
The inner splitting and subsequent partial prominence eruption is also observed by
\citet{shen12}. \citet{gil01} interpreted an active prominence with the process of vertical reconnection between an
inverse-polarity flux rope and an underlying magnetic arcade. \citet{liu08b} reported a partial filament eruption characterised
by a quasi-static, slow phase and a rapid kinking phase showing a bifurcation of the filament. The separation of the filament,
the extreme-ultravoilet (EUV) brightening at the separation location, and the surviving sigmoidal structure provide convincing
evidences that magnetic reconnection occurs within the body of filament \citep{tri13}. \citet{gib06a,gib06b} carried out
three-dimensional (3D) numerical simulations to model the partial expulsion of a MFR. After multiple reconnections
at current sheets that form during the eruption, the rope breaks in an upper, escaping rope and a lower, surviving rope. The
``partially-expelled flux rope'' (PEFR) model has been justified observationally \citep{tri09}. \citet{tri06} observed a distinct
coronal downflow following a curved path at the speed of $<$150 km s$^{-1}$ during a CME-associated prominence eruption.
Their observation provides support for the pinching off of the field lines drawn-out by the erupting prominences and the
contraction of the arcade formed by the reconnection. Similar multithermal downflow at the speed of $\sim$380 km s$^{-1}$
starting at the cusp-shaped structures where magnetic reconnection occurred inside the erupting flux rope that led to its
bifurcation was reported by \citet{tri07}. \citet{liu12} studied a flare-associated partial eruption of a double-decker filament.
\citet{cx14b}
found that a stable double-decker MFR system existed for hours prior to the eruption on 2012 July 12.
After entering the domain of instability, the high-lying MFR impulsively erupted to generate a fast CME and \textit{GOES} X1.4
class flare; while the low-lying MFR remained behind and continuously maintained the sigmoidicity of the active region (AR).
From the previous literatures, we can conclude that magnetic reconnection and the release of free energy involve in most
of the partial filament eruptions. However, the exact mechanism of partial eruptions, which is of great importance to
understanding the origin of solar eruptions and forecasting space weather, remain unclear and controversial.

In this paper, we report multiwavelength observations of a partial filament eruption and the associated CME and M6.7 flare
in NOAA AR 11283 on 2011 September 8. The AR emerged from the eastern solar limb on 2011 August 30 and lasted for 14
days. Owing to its extreme complexity, it produced a couple of giant flares and CMEs during its lifetime
\citep{feng13,dai13,jiang14,liu14,li14,ruan14}. In Section~\ref{s-data}, we describe the data analysis using observations
from the Big Bear Solar Observatory (BBSO), \textit{SOHO}, \textit{Solar Dynamics Observatory} (\textit{SDO}),
\textit{Solar Terrestrial Relation Observatory} \citep[\textit{STEREO};][]{kai05}, \textit{GOES},
\textit{Reuven Ramaty High-Energy Solar Spectroscopic Imager} \citep[\textit{RHESSI};][]{lin02}, and \textit{WIND}. Results and
discussions are presented in Section~\ref{s-result} and Section~\ref{s-disc}. Finally, we draw our conclusion in Section~\ref{s-sum}.

\section{Instruments and data analysis} \label{s-data}

\subsection{BBSO and \textit{SOHO} observations} \label{s-ha}
On September 8, the dark filament residing in the AR was most clearly observed at H$\alpha$ line center ($\sim$6563 {\AA})
by the ground-based telescope in BBSO. During 15:30$-$16:30 UT, the filament rose and split into two parts. The major part
lifted away and returned to the solar surface, while the runaway part separated from and escaped the major part, resulting
in a very faint CME recorded by the \textit{SOHO} Large Angle Spectroscopic Coronagraph \citep[LASCO;][]{bru95}
CME catalog\footnote{http://cdaw.gsfc.nasa.gov/CME\_list/}. The white light (WL) images observed by the LASCO/C2
with field-of-view (FOV) of 2$-$6 solar radii ($R_{\sun}$) were calibrated using the \textit{c2\_calibrate.pro} in the
\textit{Solar Software} (\textit{SSW}).

\subsection{\textit{SDO} observations} \label{s-euv}
The partial filament eruption was clearly observed by the Atmospheric Imaging Assembly \citep[AIA;][]{lem12}
aboard \textit{SDO} with high cadences and resolutions. There are seven EUV filters (94, 131, 171, 193, 211, 304,
and 335 {\AA}) and two UV filters (1600 {\AA} and 1700 {\AA}) aboard AIA to achieve a wide temperature coverage
($4.5\le \log T \le7.5$). The AIA level\_1 fits data were calibrated using the standard program \textit{aia\_prep.pro}.
The images observed in different wavelengths were coaligned carefully using the cross-correlation method.

To investigate the 3D magnetic configurations before and after the eruption, we employed the line-of-sight (LOS)
and vector magnetograms from the Helioseismic and Magnetic Imager \citep[HMI;][]{sch12} aboard \textit{SDO}. The
180$^{\circ}$ ambiguity of the transverse field was removed by assuming that the field changes smoothly at the
photosphere \citep{guo13}. We also performed magnetic potential field and non-linear force free field (NLFFF)
extrapolations using the
optimization method as proposed by \citet{wht00} and as implemented by \citet{wig04}. The FOV for extrapolation was
558$\farcs$5$\times$466$\farcs$2 to cover the whole AR and make sure the magnetic flux was balanced, and the data
were binned by 2$\times$2 so that the resolution became 2$\arcsec$.

\subsection{\textit{STEREO} and \textit{WIND} observations}
The eruption was also captured from different perspectives
by the Extreme-Ultraviolet Imager (EUVI) and COR1\footnote{http://cor1.gsfc.nasa.gov/catalog/cme/2011/}
coronagraph of the Sun Earth Connection Coronal and Heliospheric Investigation \citep[SECCHI;][]{how08} instrument
aboard the ahead satellite (\textit{STA} hereafter) and behind satellite (\textit{STB} hereafter) of \textit{STEREO}.
The COR1 has a smaller FOV of 1.3$-$4.0 $R_{\sun}$ compared with LASCO/C2, which is favorable for the
detection of early propagation of CMEs. On September 8, the twin satellites
(\textit{STA} and \textit{STB}) had separation angles of 103$^{\circ}$ and 95$^{\circ}$ with the Earth.

The presence of open magnetic field lines within the AR was confirmed indirectly by the evidence of type \Rmnum{3}
burst in the radio dynamic spectra. The spectra were obtained by the S/WAVES \citep{bou08} on board \textit{STEREO}
and the WAVES instrument \citep{bou95} on board the \textit{WIND} spacecraft. The frequency of S/WAVES ranges from 2.5
kHz to 16.025 MHz. The WAVES has two radio detectors: RAD1 (0.02$-$1.04 MHz) and RAD2 (1.075$-$13.825 MHz).

\subsection{\textit{GOES} and \textit{RHESSI} observations} \label{s-xray}
The accompanying M6.7 flare was obviously identified in the \textit{GOES} soft X-ray (SXR) light curves in 0.5$-$4.0 {\AA}
and 1$-$8 {\AA}. To figure out where the accelerated nonthermal particles precipitate, we also made hard X-ray (HXR) images and
light curves at different energy bands (3$-$6, 6$-$12, 12$-$25, 25$-$50, and 50$-$100 keV) using the observations of \textit{RHESSI}.
The HXR images were generated using the CLEAN method with integration time of 10 s.
The observing parameters are summarized in Table~\ref{tbl-1}.

\section{Results} \label{s-result}
\begin{figure}
\epsscale{.50}
\plotone{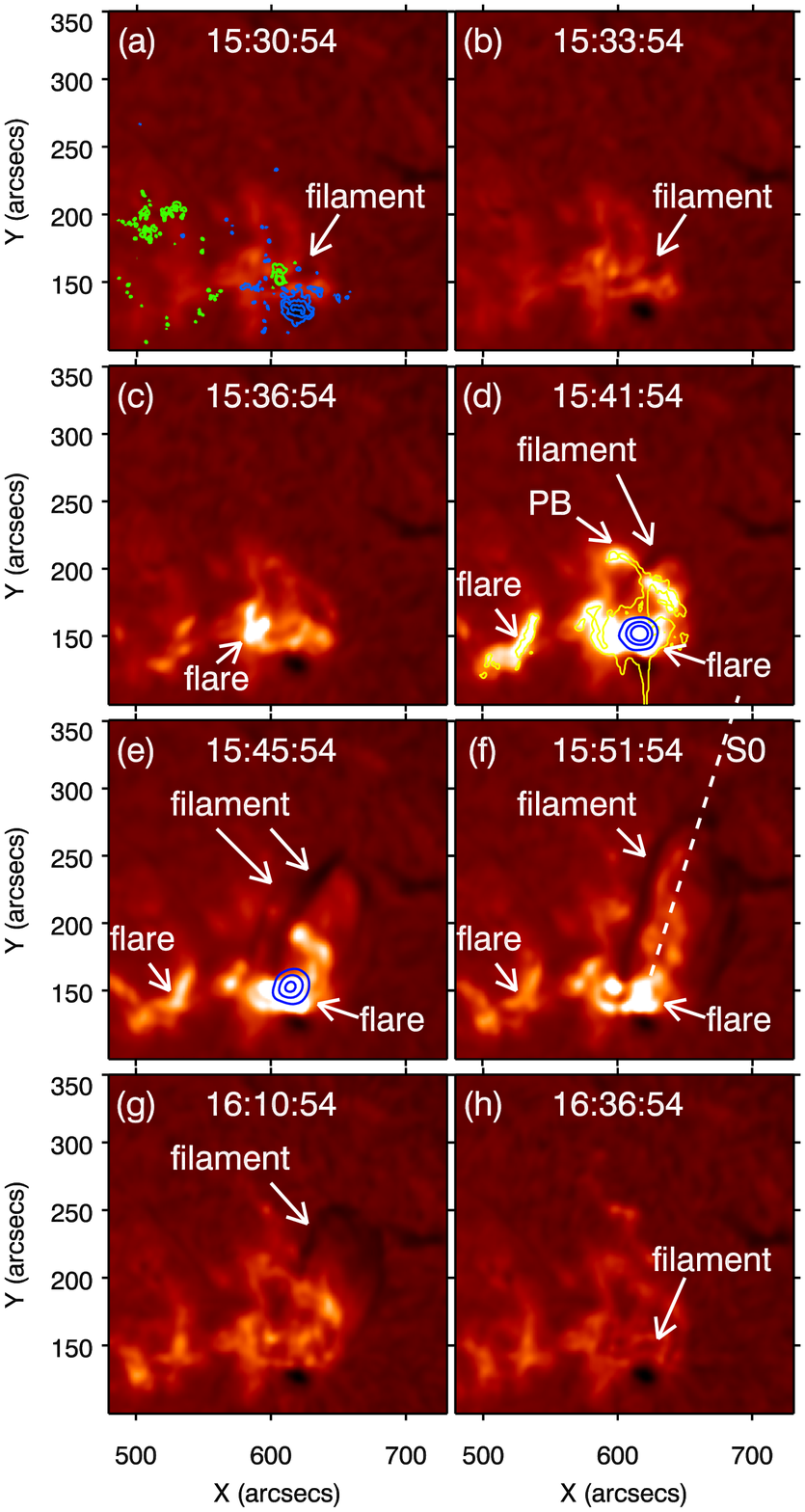}
\caption{(a)$-$(h) Eight snapshots of the H$\alpha$ images observed by BBSO.
The white arrows in panels (a)$-$(e) point to the dark filament, bright flare, and
point-like brightening (PB). The solid blue lines in panels (d) and (e) denote the
intensity contours of the HXR emission at 12$-$25 keV. The solid yellow lines in
panel (d) denote the intensity contours of the AIA 1600 {\AA} intensity at the same
time. The dashed line labeled with ``S0'' in panel (f)
is used for investigating the evolution of the major part of the filament whose
time-slice diagram is displayed in Figure~\ref{fig8}(a).
\label{fig1}}
(Animations of this figure are available in the online journal.)
\end{figure}

\begin{figure}
\centerline{\hspace*{0.01\textwidth}
            \includegraphics[width=0.45\textwidth,clip=]{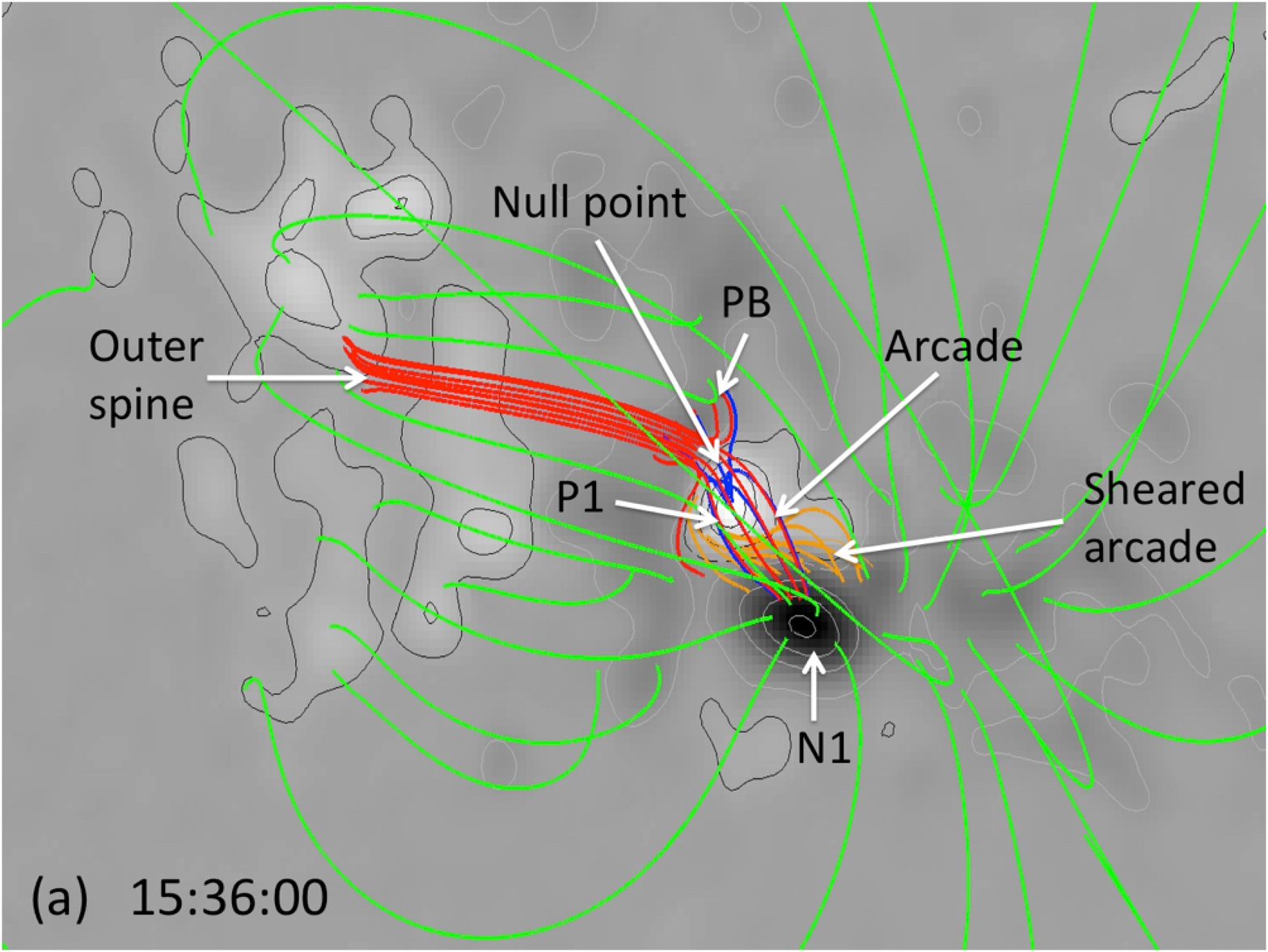}
            \hspace*{0.01\textwidth}
            \includegraphics[width=0.45\textwidth,clip=]{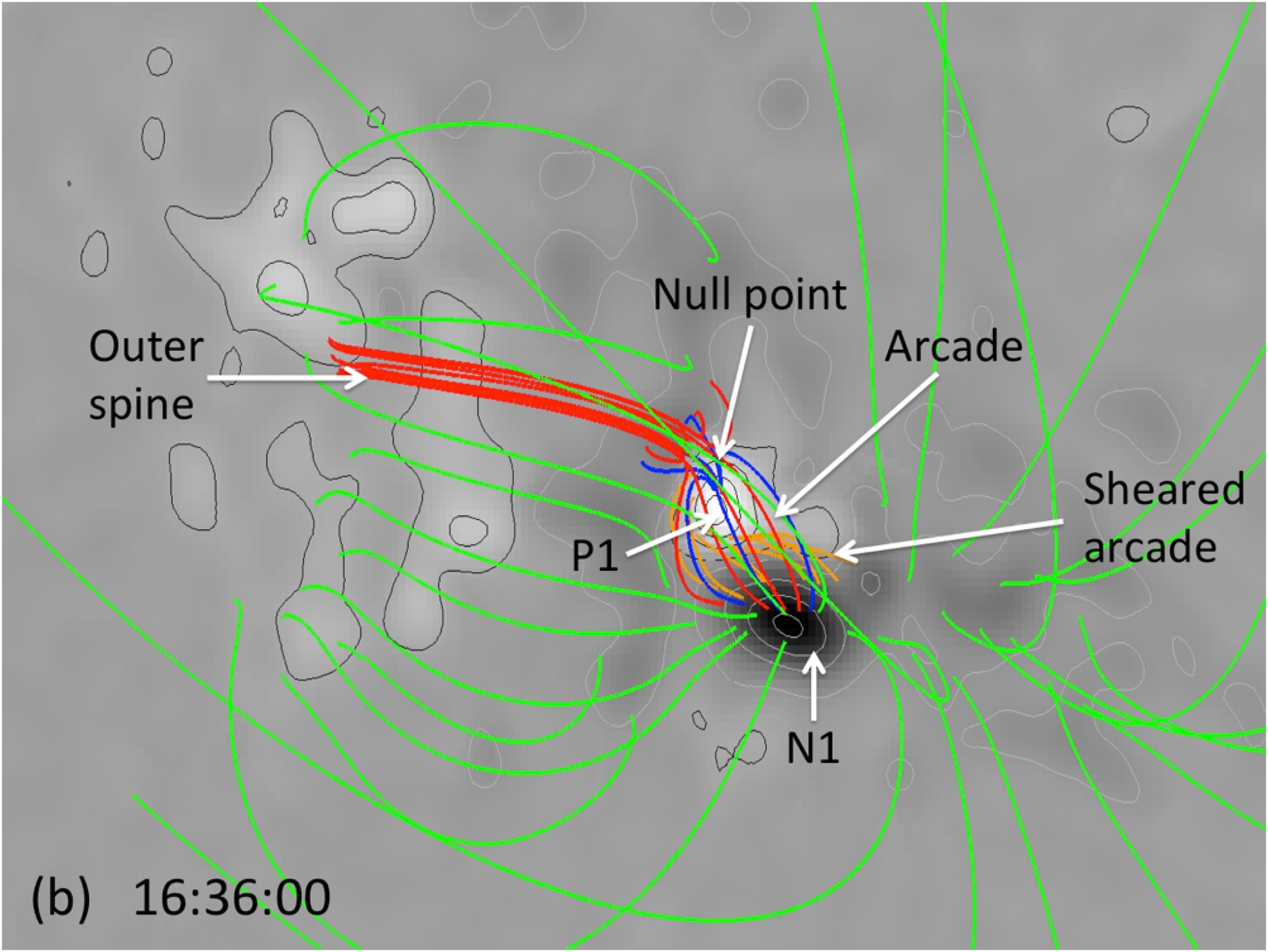}
           }

  \centerline{\hspace*{0.01\textwidth}
              \includegraphics[width=0.45\textwidth,clip=]{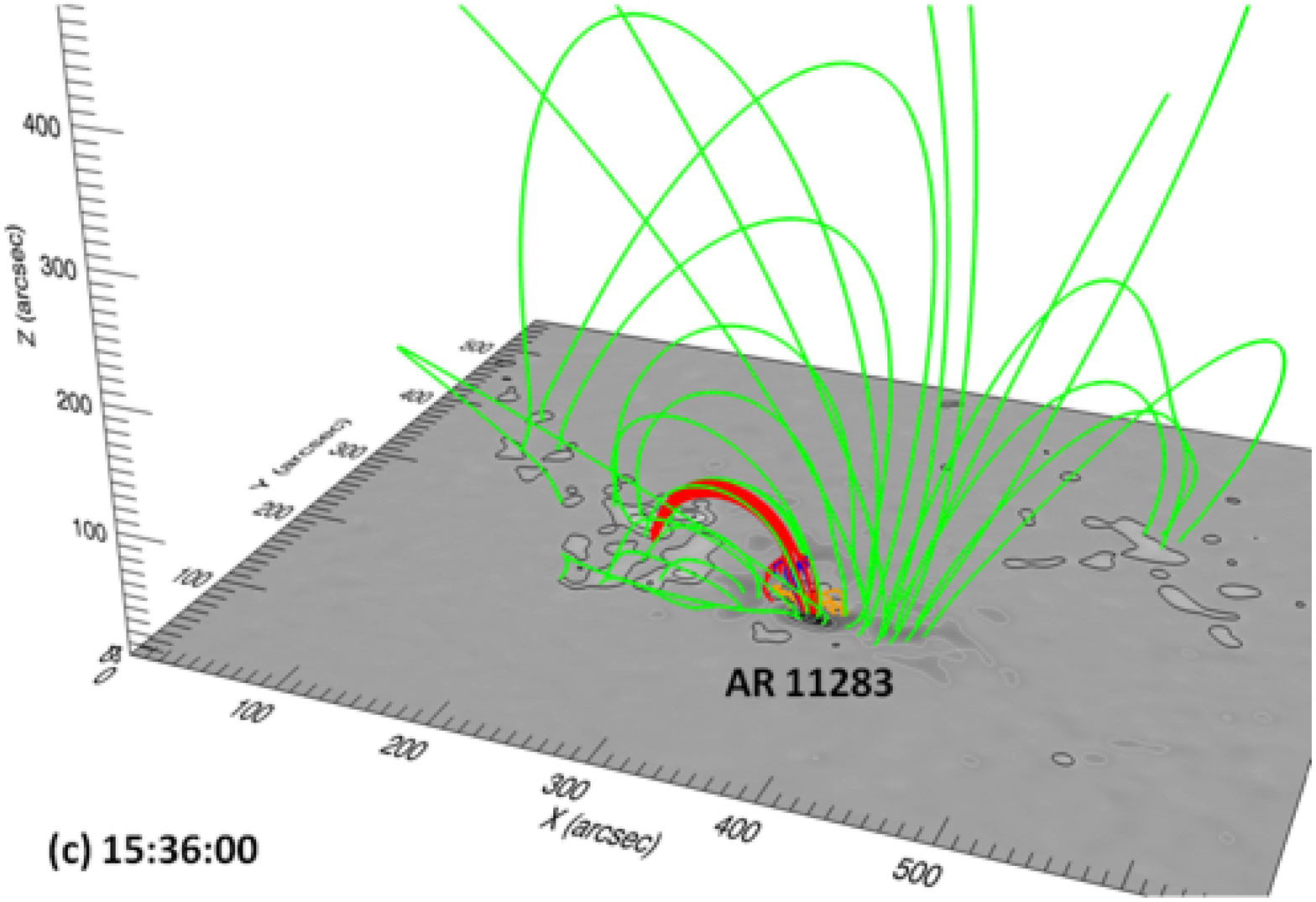}
              \hspace*{0.01\textwidth}
              \includegraphics[width=0.45\textwidth,clip=]{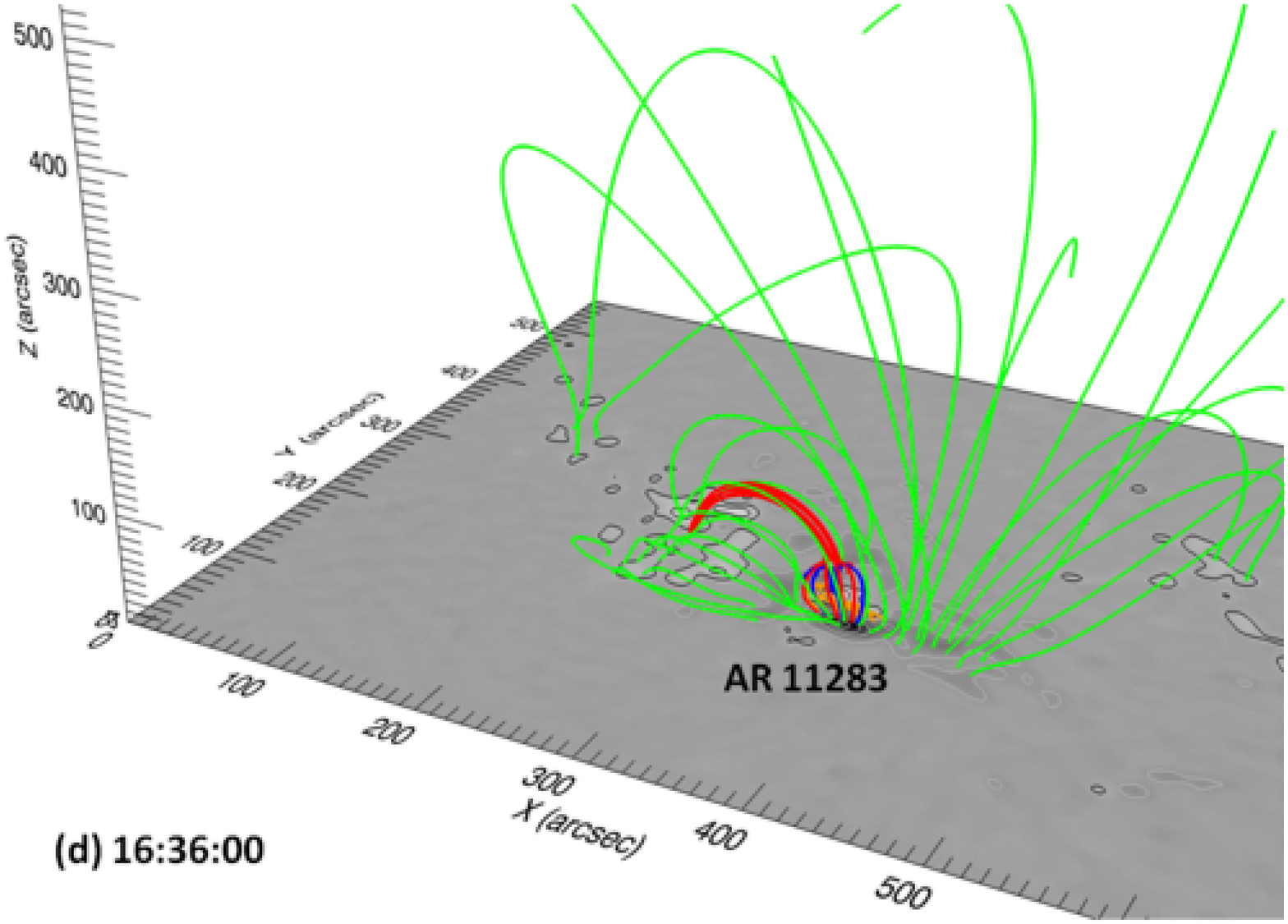}
             }
\caption{(a)$-$(b) Top-view of the 3D magnetic configuration at 15:36 UT and 16:36 UT
with the LOS magnetograms located at the bottom boundary of the boxes. The arrows
point to the null point, outer spine, sheared arcade supporting the filament, arcade (or separatrix)
above the filament, PB, positive polarity P1, and negative polarity N1.
(c)$-$(d) Side-view of the 3D magnetic configuration at 15:36 UT and 16:36 UT.
The green lines represent the normal magnetic field lines. The red/blue lines represent
field lines near the outer/inner spine and the fan surface. The orange lines represent the field
lines of the sheared arcade.
\label{fig2}}
\end{figure}

Figure~\ref{fig1} shows eight snapshots of the H$\alpha$ images to illustrate the whole evolution of the filament
(see also the online movie Animation1.mpg). Figure~\ref{fig1}(a) displays the H$\alpha$ image at 15:30:54 UT
before eruption. It is overlaid with the contours of the LOS magnetic field, where green (blue) lines stand for positive
(negative) polarities. The dark filament that is $\sim$39 Mm long resides along the polarity inversion line (PIL).
The top panels of
Figure~\ref{fig2} demonstrate the top-view of the 3D magnetic configuration above the AR at the beginning and after
eruption, with the LOS magnetograms located at the bottom boundary. Using the same method described in \citet{zhang12c},
we found a magnetic null point and the corresponding spine and separatrix surface. The normal magnetic field
lines are denoted by green lines. The magnetic field lines around the outer/inner spine and the separatrix surface
(or arcade) are represented by red/blue lines. Beneath the null point, the sheared arcades supporting the filament are
represented by orange lines. The spine is rooted in the positive polarity (P1) that is surrounded by the negative polarities
(N1 and PB). It extends in the northeast direction and connects the null point with a remote place on the solar surface.
Such magnetic configuration is quite similar to those reported by \citet{sun12b}, \citet{jiang13}, and \citet{man14}.

\begin{figure}
\epsscale{.50}
\plotone{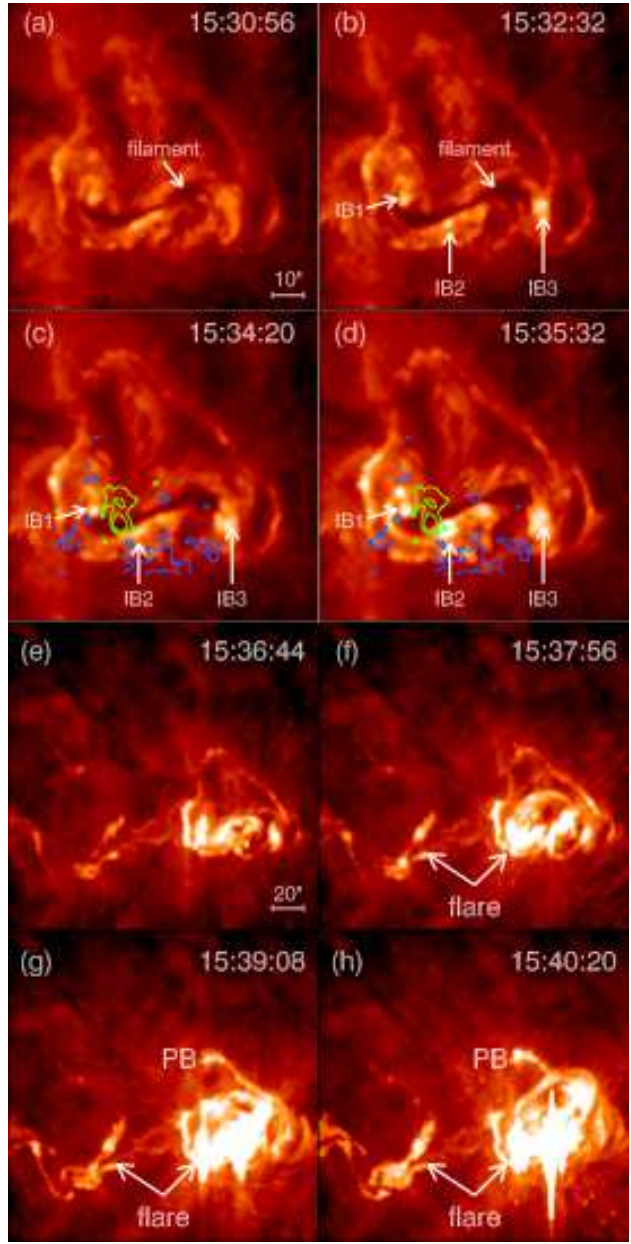}
\caption{(a)$-$(h) Eight snapshots of the AIA 304 {\AA} images. The FOV of the
lower four panels is larger than that of the upper four panels. The arrows in panel (a)-(b)
point to the dark filament. The arrows in panels (b)-(d) point to the initial brightenings
(IB1, IB2, and IB3) at the two ends and center of the filament. The arrows in panel
(f)-(h) point to the flare and PB.
\label{fig3}}
(Animations of this figure are available in the online journal.)
\end{figure}

\begin{figure}
\epsscale{.90}
\plotone{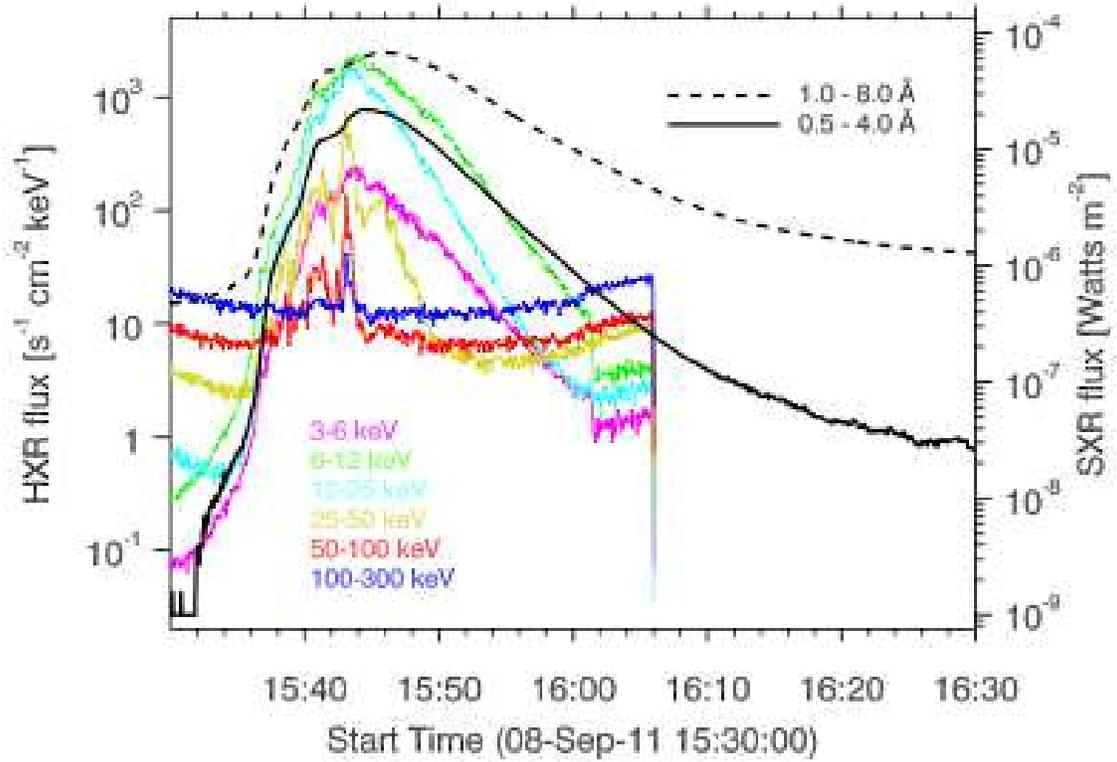}
\caption{SXR and HXR light curves
of the flare associated with the filament eruption. The black solid and dashed lines represent
the SXR light curves in 0.5$-$4.0 {\AA} and 1$-$8 {\AA}. The colored lines denote the HXR light
curves at different energy bands.
\label{fig4}}
\end{figure}

As time goes on, the filament rose and expanded slowly (Figure~\ref{fig1}(b)). The initiation process is clearly
revealed by the AIA 304 {\AA} observation (see the online movie Animation2.mpg). Figure~\ref{fig3} shows eight
snapshots of the 304 {\AA} images. Initial brigtenings (IB1, IB2, and IB3) appeared near the ends and center of the
sigmoidal filament, implying that magnetic reconnection took place and the filament got unstable (Figure~\ref{fig3}(b)-(d)).
Such initial brightenings were evident in all the EUV wavelengths. With the intensities of the brigtenings increasing, the dark
filament rose and expanded slowly, squeezing the overlying arcade field lines. Null-point magnetic reconnection might be
triggered when the filament reached the initial height of the null point ($\sim$15 Mm), leading to impulsive brightenings in
H$\alpha$ (Figure~\ref{fig1}(c)-(d)) and EUV (Figure~\ref{fig3}(e)-(h)) wavelengths and increases in SXR and HXR
fluxes (Figure~\ref{fig4}). The M6.7 flare entered the impulsive phase. The bright and compact flare kernel pointed by the
white arrow in Figure~\ref{fig1}(c) extended first westward and then northward, forming a quasi-circular ribbon at
$\sim$15:42 UT (Figure~\ref{fig1}(d)), with the intensity contours of the HXR emissions at 12$-$25 keV superposed.
There was only one HXR source associated with the flare, and the source was located along the flare ribbon with the
strongest H$\alpha$ emission, which is compatible with the fact that the footpoint HXR emissions come from the nonthermal
bremsstrahlung of the accelerated high-energy electrons after penetrating into the chromosphere. The flare demonstrates
itself not only around the filament but also at the point-like brightening (PB hereafter) and the V-shape ribbon to the left of
the quasi-circular ribbon. Since the separatrix surface
intersects with the photosphere at PB to the north and the outer spine intersects with the photosphere to the east
(Figure~\ref{fig2}(a)), it is believed that nonthermal electrons accelerated by the null-point magnetic reconnection
penetrated into the lower atmosphere not only at the quasi-circular ribbon, but also at PB and the V-shape ribbon.

Figure~\ref{fig4} shows the SXR (black solid and dashed lines) and HXR (colored solid lines) light curves of the flare.
The SXR fluxes started to rise rapidly at $\sim$15:32 UT and peaked at 15:45:53 UT for 1$-$8 {\AA} and 15:44:21 UT
for 0.5$-$4.0 {\AA}. The HXR fluxes below 25 keV varied smoothly like the SXR fluxes, except for earlier peak times at
$\sim$15:43:10 UT. The HXR fluxes above 25 keV, however, experienced two small peaks that imply precursor release
of magnetic energy and particle acceleration at $\sim$15:38:36 UT and $\sim$15:41:24 UT and a major peak at
$\sim$15:43:10 UT. The time delay between the SXR and HXR peak times implies the possible Neupert effect for this
event \citep{ning10}. The main phase of the flare sustained until $\sim$17:00 UT, indicating that the flare is a long-duration
event.

\begin{figure}
\epsscale{.50}
\plotone{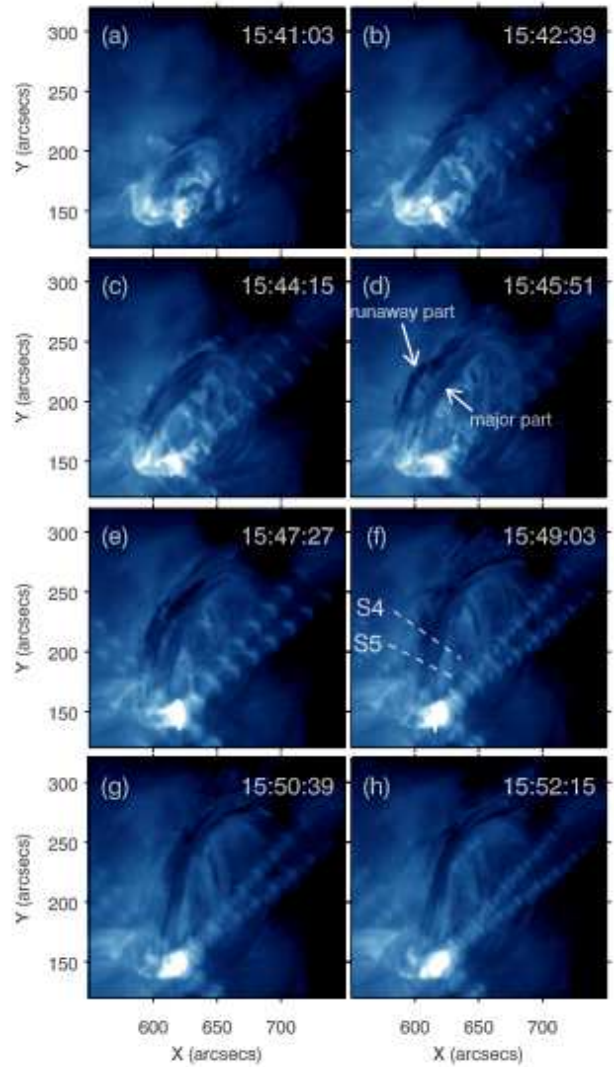}
\caption{(a)$-$(h) Eight snapshots of the AIA 335 {\AA} images. The filament split into
the runaway part and major part at $\sim$15:46 UT (panel (d)). The white dashed lines
in panel (f) labeled with ``S4'' and ``S5'' are used for investigating the rotation of the
filament at the eastern leg. The time-slice diagrams of the two slices are drawn in
Figure~\ref{fig6}.
\label{fig5}}
(Animations of this figure are available in the online journal.)
\end{figure}

\begin{figure}
\epsscale{.80}
\plotone{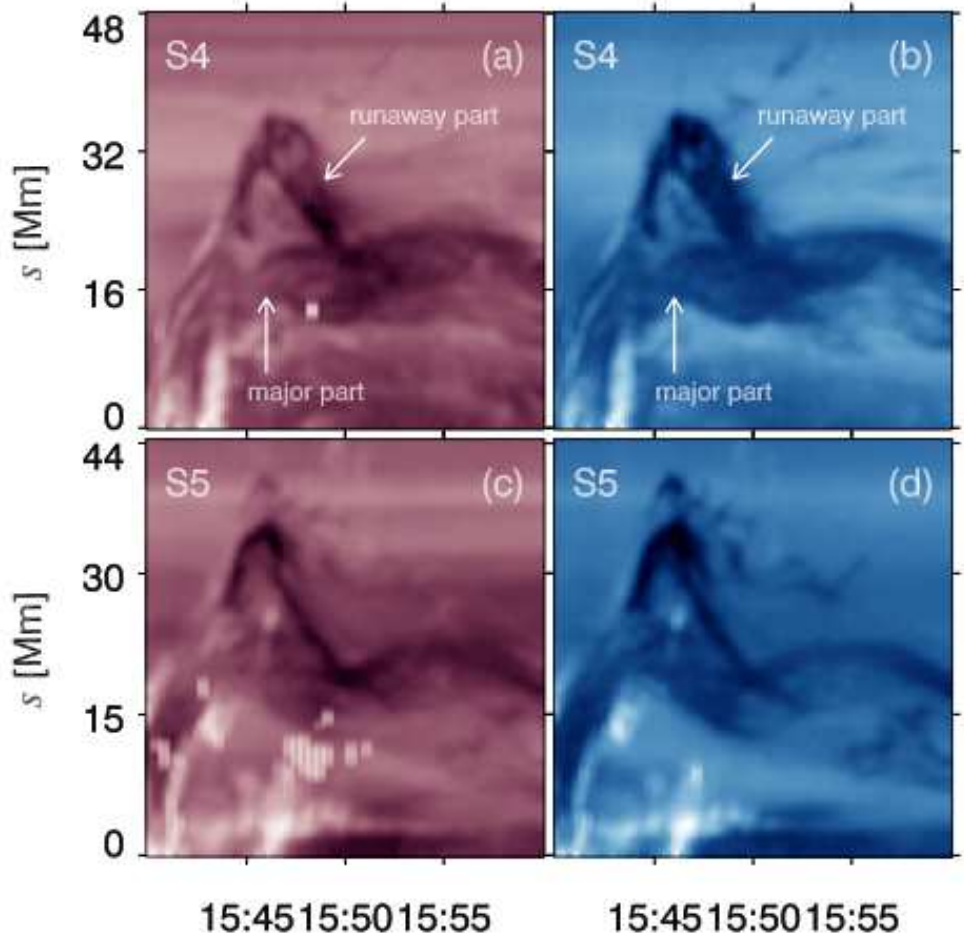}
\caption{Time-slice diagrams of S4 (upper panels) and S5 (lower panels) in
211 {\AA} (left panels) and 335  {\AA} (right panels), showing the rotation of
the runaway part of the filament around the major part for $\sim$1 turn.
\label{fig6}}
\end{figure}

During the flare, the filament continued to rise and split into two branches at the eastern leg around 15:46 UT
(Figure~\ref{fig1}(e)), the right of which is thicker and darker than the left one. Such a process is most clearly
revealed by the AIA 335 {\AA} observation (see the online movie Animation3.mpg). Figure~\ref{fig5} displays eight
snapshots of the 335 {\AA} images. It is seen that the dark filament broadened
from $\sim$15:42:30 UT and completely split into two branches around 15:45:51 UT. We define the left and right branches
as the runaway part and major part of the filament. The two interwinding parts also underwent rotation (panels (d)-(h)).
Meanwhile, the plasma of the runaway part moved in the northwest direction and escaped. To illustrate the rotation,
we derived the time-slice diagrams of the two slices (S4 and S5 in panel (f)) that are plotted in Figure~\ref{fig6}. The upper
(lower) panels represent the diagrams of S4 (S5), and the left (right) panels represent the diagrams for 211 {\AA} (335 {\AA}).
$s=0$ in the diagrams stands for the southwest endpoints of the slices. The filament began to split into two parts around
15:42:30 UT, with the runaway part rotating round the eastern leg of the major part for $\sim$1 turn until $\sim$15:55 UT.

During the eruption, the runaway branch of the filament disappeared (Figure~\ref{fig1}(f)). The major part, however,
fell back to the solar surface after reaching the maximum height around 15:51 UT, suggesting that the eruption of the major
part of the filament was failed. The remaining filament after the flare was evident in the H$\alpha$ image
(Figure~\ref{fig1}(h)). NLFFF modelling shows that the magnetic topology was analogous to that before the flare, with the
height of the null point slightly increased by 0.4 Mm (Figure~\ref{fig2}(b)).

\begin{figure}
\epsscale{.50}
\plotone{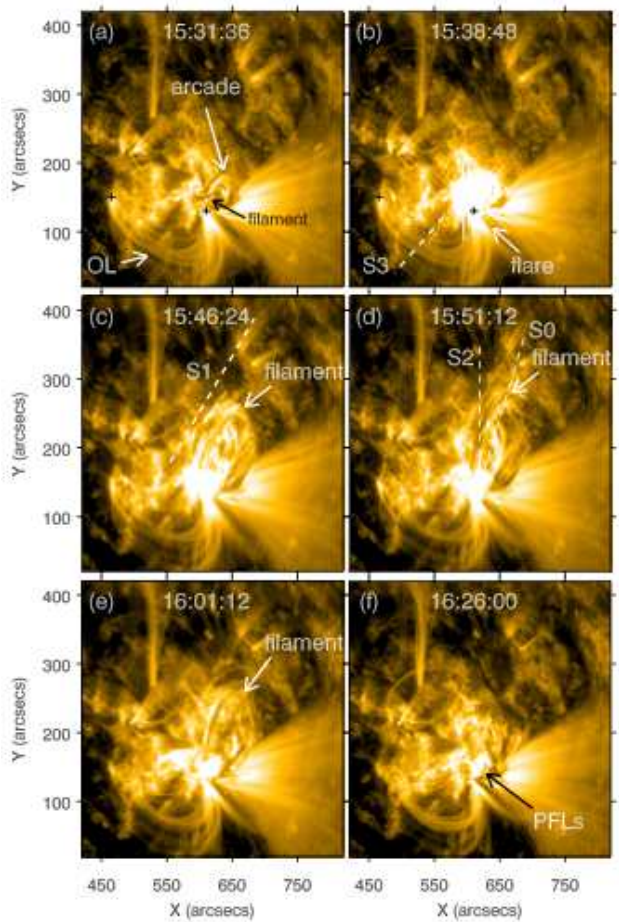}
\caption{(a)$-$(f) Six snapshots of the AIA 171 {\AA} images. The arrows in panels
(a)$-$(e) point to the oscillating coronal loop (``OL''), arcade, filament, and flare.
The dashed lines labeled with ``S3'' in panel (b) and ``S1'' in panel (c) are used for
investigating the temporal evolutions of the OL and the escaping material, respectively.
The dashed line labeled with ``S2'' that passes through the runaway and major parts
of the filament in panel (d) is used for studying the bifurcation of the filament. The
time-slice diagrams of S1, S2, and S3 are displayed in Figure~\ref{fig8}(b)-(d).
The dashed line labeled with ``S0'' in panel (d) has the same meaning as that in
Figure~\ref{fig1}(f). The black crosses in panels (a) and (b) represent the
footpoints of the OL. The arrow in panel (f) point to the post-flare loops (PFLs).
\label{fig7}}
(Animations of this figure are available in the online journal.)
\end{figure}

Figure~\ref{fig7} shows six snapshots of the 171 {\AA} images. The rising and expanding filament triggered the M-class flare
and the kink-mode oscillation of the adjacent large-scale coronal loops within the same AR (see the online movie Animation4.mpg).
With the filament increasing in height, part of its material was ejected in the northwest direction represented by ``S1'' in panel
(c). After reaching the maximum height at $\sim$15:51:12 UT, the major part of the filament returned to the solar surface. The
bright cusp-like post-flare loops (PFLs) in the main phase of the flare are clearly observed in all the EUV filters, see also
Figure~\ref{fig7}(f).

\begin{figure}
\epsscale{.40}
\plotone{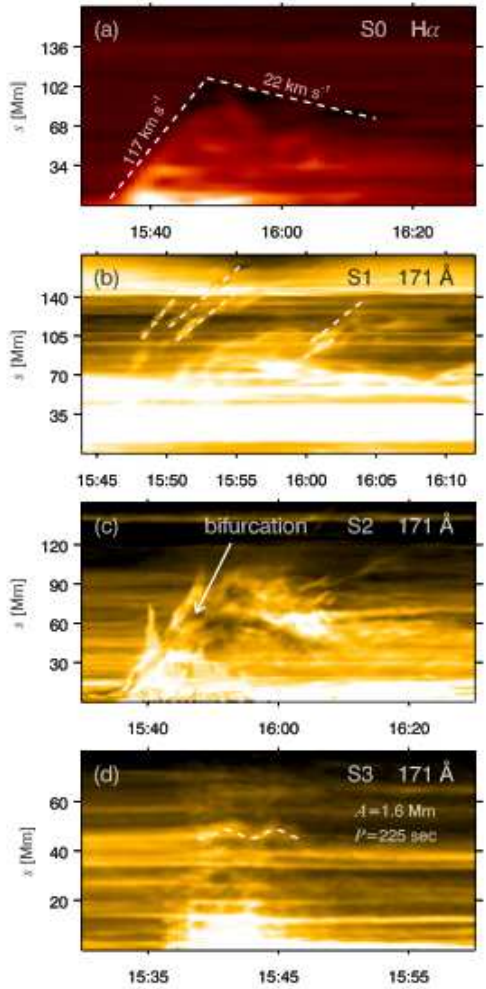}
\caption{(a) Time-slice diagram of S0 in H$\alpha$ wavelength. The slopes of the
dashed lines denote the rising ($\sim$117 km s$^{-1}$) and falling ($\sim$22 km s$^{-1}$)
speeds of the major part of the filament. (b) Time-slice diagram of S1 in 171 {\AA}.
The slopes of the dashed lines stand for the apparent velocities of the escaping material
along S1, ranging from 125 km s$^{-1}$ to 255 km s$^{-1}$. (c) Time-slice diagram of S2
in 171 {\AA}. The white arrow point to the time of bifurcation of the two parts of the filament.
(d) Time-slice diagram of S3 in 171 {\AA}. The dashed line represents the kink oscillation
of the coronal loop. The values of amplitude ($A$) and period ($P$) are presented.
\label{fig8}}
\end{figure}

\begin{figure}
\epsscale{.60}
\plotone{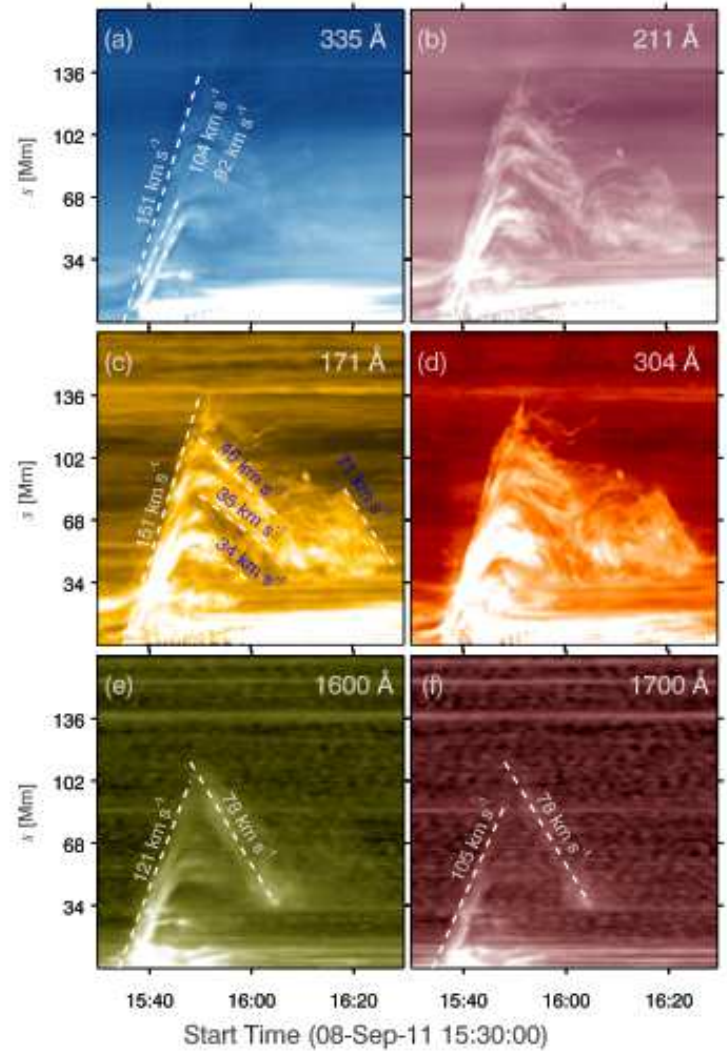}
\caption{(a)$-$(d) Time-slice diagrams of S0 in four EUV filters of AIA
(335, 211, 171, and 304 {\AA}). The slopes of the dashed lines in panel (a)
stand for the rising velocities of the filament, being 151, 104, and 92 km s$^{-1}$. The
slopes of the dashed lines in panel (c) represent the rising (151 km s$^{-1}$) and falling
(34, 35, 46, and 71 km s$^{-1}$) speeds of the filament.
(e)$-$(f) Time-slice diagrams of S0 in two UV filters of AIA (1600 {\AA} and 1700 {\AA}).
The slopes of the dashed lines signify the rising (121 and 105 km s$^{-1}$) and falling
(78 km s$^{-1}$) speeds of the filament.
\label{fig9}}
\end{figure}

To illustrate the eruption and loop oscillation more clearly, we extracted four slices. The first slice, S0 in Figure~\ref{fig1}(f)
and Figure~\ref{fig7}(d), is 170 Mm in length. It starts from the flare site and passes through the apex of the major part of the
filament. The time-slice diagram of S0 in H$\alpha$ is displayed in Figure~\ref{fig8}(a). The filament started to rise rapidly at
$\sim$15:34:30 UT with a constant speed of $\sim$117 km s$^{-1}$. After reaching the peak height ($z_{max}$) of $\sim$115
Mm at $\sim$15:51 UT, it fell back to the solar surface in a bumpy way until $\sim$16:30 UT. Using a linear fitting, we derived
the average falling speed ($\sim$22 km s$^{-1}$) of the filament in H$\alpha$ wavelength. The time-slice diagram of S0 in UV
and EUV passbands are presented in Figure~\ref{fig9}. We selected two relatively hot filters (335 {\AA} and 211 {\AA}
in the top panels), two warm filters (171 {\AA} and 304 {\AA} in the middle panels), and two cool filters (1600 {\AA} and 1700
{\AA} in the bottom panels), respectively. Similar to the time-slice diagram in H$\alpha$ (Figure~\ref{fig8}(a)), the filament rose
at apparent speeds of 92$-$151 km s$^{-1}$ before felling back in an oscillatory way at the speeds of 34$-$46 km s$^{-1}$
during 15:51$-$16:10 UT and $\sim$71 km s$^{-1}$ during 16:18$-$16:30 UT (Figure~\ref{fig9}(a)-(d)). The falling speeds in UV
wavelengths are $\sim$78 km s$^{-1}$ during 15:51$-$16:10 UT (Figure~\ref{fig9}(e)-(f)). The times when the major part of the
filament reached maximum height in UV and EUV passbands, $\sim$15:51 UT, are consistent with that in H$\alpha$. The later
falling phase during 16:18$-$16:30 UT is most obvious in the warm filters. The downflow of the surviving filament in an
oscillatory way was also observed during a sympathetic filament eruption \citep{shen12}.

Owing to the lower time cadence of BBSO than AIA, the escaping process of the runaway part of the filament in
Figure~\ref{fig1}(e)-(f) was detected by AIA. We extracted another slice S1 that is 177 Mm in length along the direction of
ejection (Figure~\ref{fig7}(c)). $s=0$ Mm and $s=177$ Mm represent the southeast and northwest endpoints of the slice.
The time-slice diagram of S1 in 171 {\AA} is displayed in Figure~\ref{fig8}(b). Contrary to the major part, the runaway part
of the filament escaped successfully from the corona at the speeds of 125$-$255 km s$^{-1}$ without returning to the solar
surface. The intermittent runaway process during 15:45$-$16:05 UT was obviously observed in most of the EUV filters.
We extracted another slice S2 that also starts from the flare site and passes through both parts of the filament
(Figure~\ref{fig7}(d)). The time-slice diagram of S2 in 171 {\AA} is drawn in Figure~\ref{fig8}(c). As expected, the diagram
features the bifurcation of the filament as pointed by the white arrow, i.e., the runaway part escaped forever while the
major part moved on after bifurcation and finally fell back.

The eruption of the filament triggered transverse kink oscillation of the adjacent coronal loops (OL in Figure~\ref{fig7}(a)).
The direction of oscillation is perpendicular to the initial loop plane (see the online movie Animation4.mpg).
We extracted another slice S3 that is 80 Mm in length across the oscillating loops (Figure~\ref{fig7}(b)). $s=0$ Mm and
$s=80$ Mm represent the northwest and southeast endpoints of the slice. The time-slice diagram of S3 in 171 {\AA} is shown
in Figure~\ref{fig8}(d), where the oscillation pattern during 15:38$-$15:47 UT is evidently demonstrated. The OL moved away
from the flare site during 15:38$-$15:41 UT before returning to the initial position and oscillating back and forth for $\sim$2
cycles. By fitting the pattern with a sinusoidal function as marked by the white dashed line, the resulting amplitude and period
of the kink oscillation were $\sim$1.6 Mm and $\sim$225 s. We also extracted several slices across the OL and derived the
time-slice diagrams, finding that the coronal loops oscillated in phase and the mode was fundamental.
The initial velocity amplitude of the oscillation was $\sim$44.7 km s$^{-1}$.
The speed of propagation of the mode $C_K=2L/P=\sqrt{2/(1+\rho_o/\rho_i)}v_A$, where $L$ is the loop length,
$P$ is the period, $v_A$ is the Alfv\'{e}n wave speed, and $\rho_i$ and $\rho_o$ are the plasma densities inside and
outside the loop \citep{nak99,nak01,whi12}. In Figure~\ref{fig7}(a), we denote the footpoints of the OL with black crosses that are
106.1 Mm away. Assuming a semi-circular shape, the length of the loop $L=166.7$ Mm and $C_{K}=1482$ km s$^{-1}$.
Using the same value of $\rho_o/\rho_{i}=0.1$, we derived $v_{A}=1100$ km s$^{-1}$. In addition, we estimated the
electron number density of the OL to be $\sim$2.5$\times10^{10}$ cm$^{-3}$ based on the results of NLFFF
extrapolation in Figure~\ref{fig2}(a). The kink-mode oscillation
of the loops was best observed in 171 {\AA}, indicating that the temperatures of loops were $\sim$0.8 MK.

\begin{figure}
\epsscale{.80}
\plotone{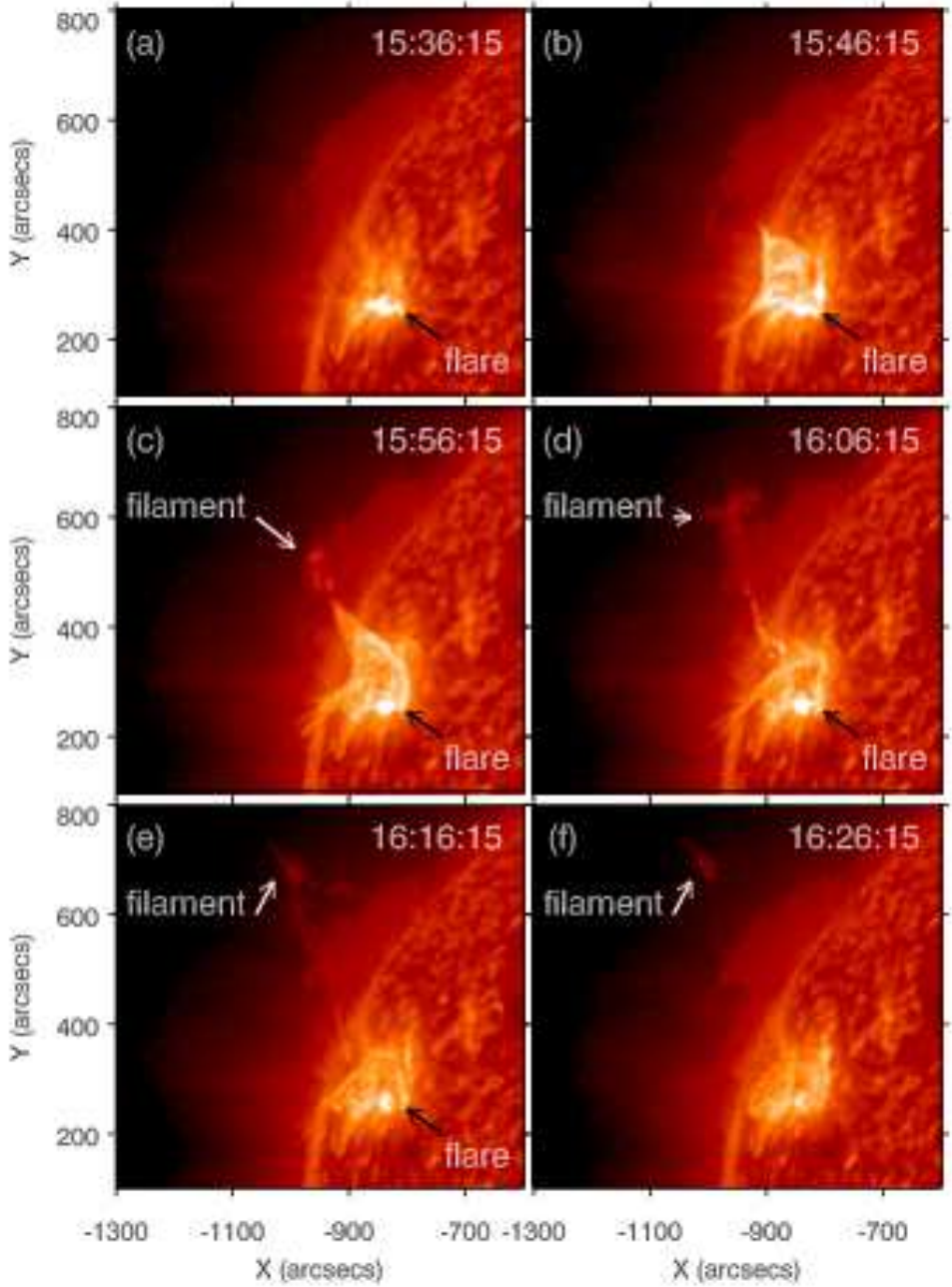}
\caption{Six snapshots of the 304 {\AA} images observed by \textit{STA}/EUVI. The white arrows
point to the escaping part of the filament in panels (c)$-$(f), while the black arrows point to the
flare in panels (a)$-$(e).
\label{fig10}}
\end{figure}

The escaping part of the filament was also clearly observed by \textit{STA}/EUVI. Figure~\ref{fig10} shows six snapshots of
the 304 {\AA} images, where the white arrows point to the escaping filament. During 15:46$-$16:30 UT, the material
moved outwards in the northeast direction without returning to the solar surface. The bright M6.7 flare pointed by the black
arrows is also quite clear.

\begin{figure}
\epsscale{.70}
\plotone{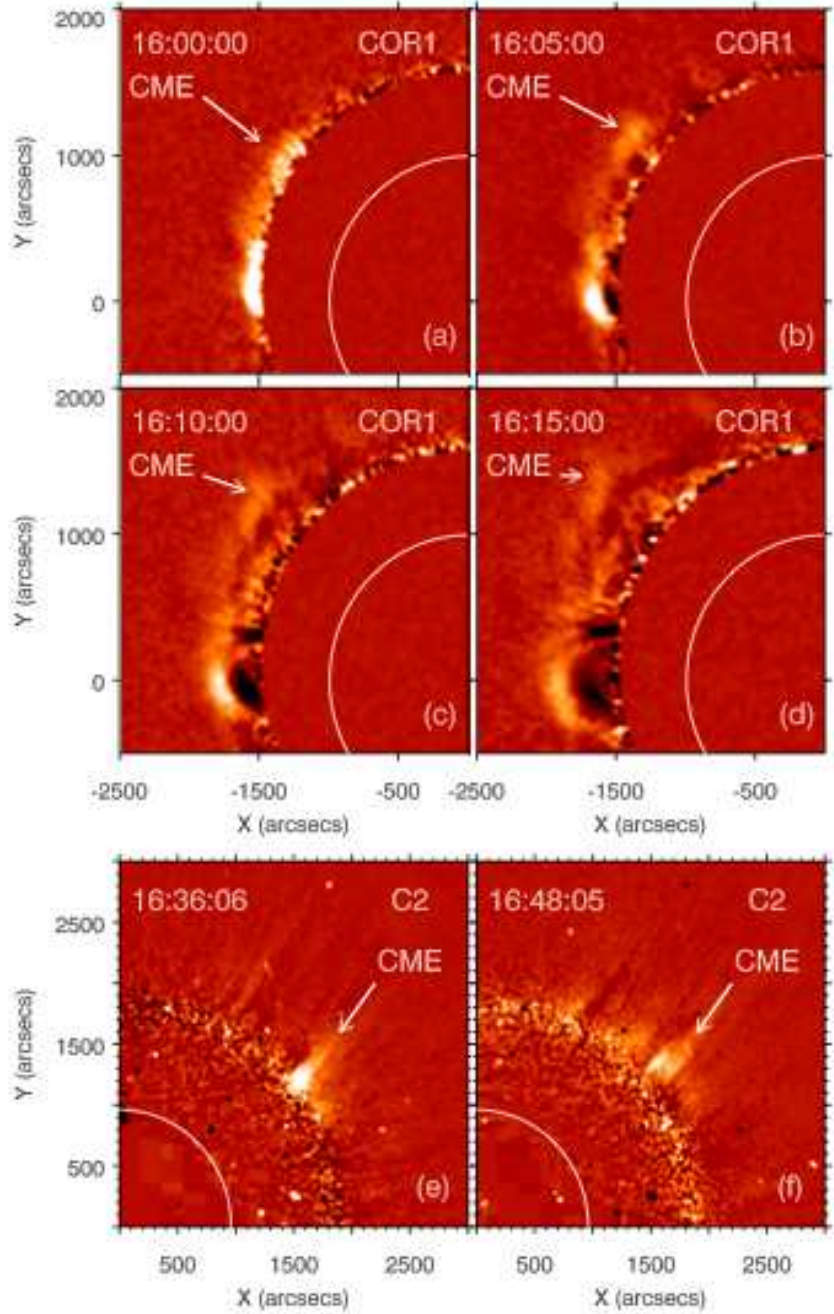}
\caption{Running-difference images of the CME observed by \textit{STA}/COR1
during 16:00$-$16:15 UT ((a)$-$(d)) and by LASCO/C2 during 16:36$-$16:48 UT
((e)$-$(f)). The white arrows point to the CME. The white arc in each panel denotes
the solar limb.
\label{fig11}}
\end{figure}

\begin{figure}
\epsscale{.90}
\plotone{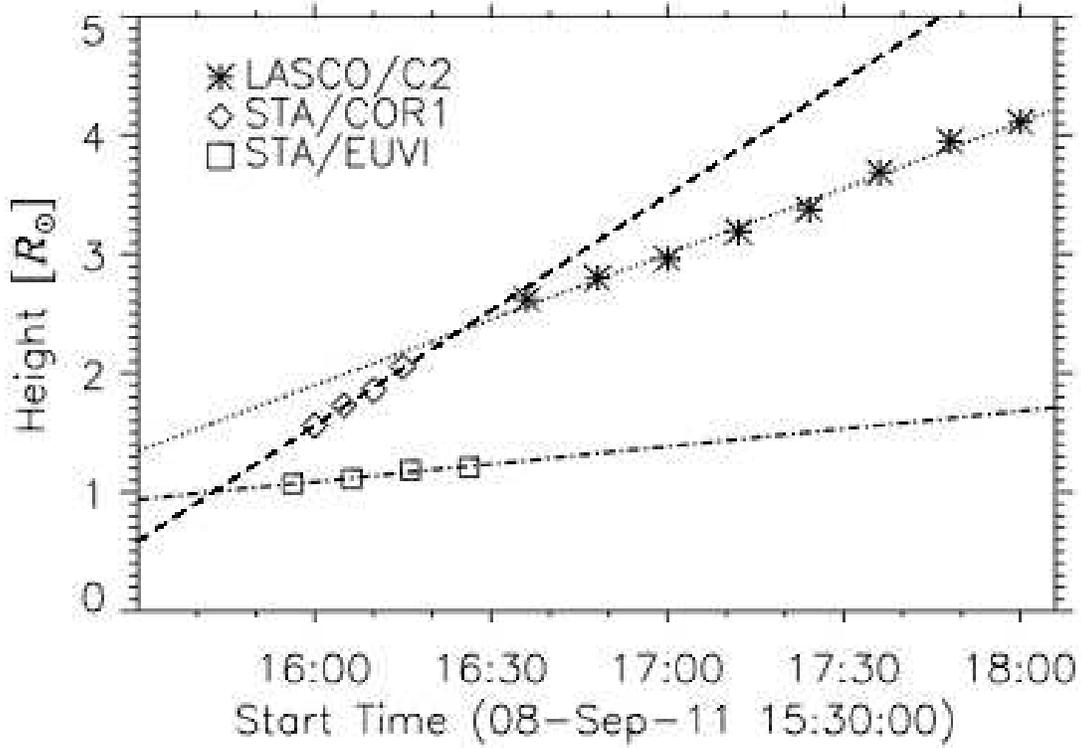}
\caption{Time-height profiles of the runaway part of the filament observed by \textit{STA}/EUVI
(\textit{boxes}), CME observed by \textit{STA}/COR1 (\textit{diamonds}), and CME observed by LASCO/C2
(\textit{stars}), respectively. The dash-dotted, dashed, and dotted lines are results of best linear
fitting whose slopes stand for the apparent propagation velocities. The height in unit of
$R_{\sun}$ signifies the heliocentric distances of the filament and CME.
\label{fig12}}
\end{figure}

The runaway part of the filament resulted in a very faint CME observed by the WL coronagraphs. Figure~\ref{fig11}(a)-(d)
show the running-difference images of \textit{STA}/COR1 during 16:00$-$16:15 UT. As pointed by the arrows, the CME first
appeared in the FOV of \textit{STA}/COR1 at $\sim$16:00 UT and propagated outwards at a nearly constant speed, with the
contrast between CME and the background decreasing as time goes on. The propagation direction of the CME is consistent
with that of the runaway filament in Figure~\ref{fig10}. Figure~\ref{fig11}(e)-(f) show the running-difference images of LASCO/C2
during 16:36$-$16:48 UT. The faint blob-like CME first appeared in the FOV of C2 at $\sim$16:36 UT and propagated in the
same direction as that of the escaping filament observed by AIA in Figure~\ref{fig7}(c). The central position angle and angular
width of the CME observed by C2 are 311$^{\circ}$ and 37$^{\circ}$. The linear velocity of the CME is $\sim$214 km s$^{-1}$.
The time-height profiles of the runaway filament observed by \textit{STA}/EUVI (\textit{boxes}) and the corresponding CME observed
by \textit{STA}/COR1 (\textit{diamonds}) and LASCO/C2 (\textit{stars}) are displayed in Figure~\ref{fig12}. The apparent propagating
velocities represented by the slopes of the lines are 60, 358, and 214 km s$^{-1}$, respectively.
Taking the projection effect into account, the start times of the filament eruption and the CME observed by LASCO/C2 and
\textit{STA}/COR1 from the lower corona ($\approx1.0 R_{\sun}$) are approximately coincident with each other. In the CDAW
catalog, the preceding and succeeding CMEs occurred at 06:12 UT and 18:36 UT on September 8. In the COR1 CME catalog,
the preceding and succeeding CMEs occurred slightly earlier at 05:45 UT and 18:05 UT on the same day, which is due to the
smaller FOV of COR1 than LASCO/C2. Therefore, the runaway part of the filament was uniquely associated with the CME
during 16:00$-$18:00 UT.

\begin{figure}
\epsscale{.70}
\plotone{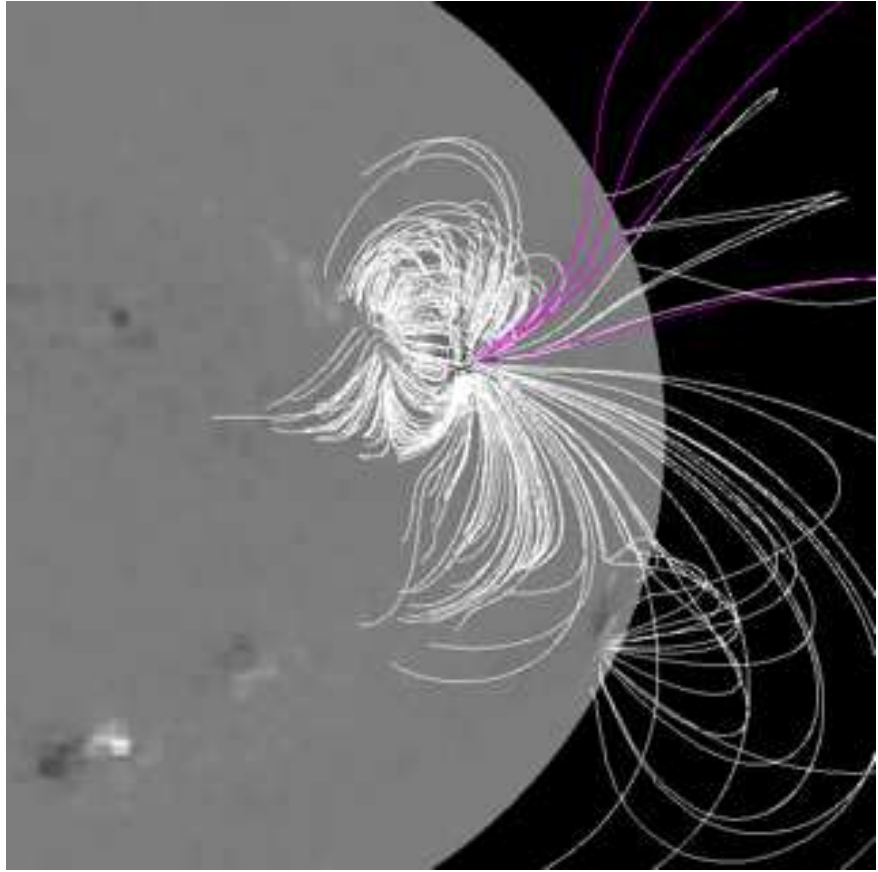}
\caption{Large-scale magnetic field lines around AR 11283 at 12:04 UT obtained by
the PFSS modelling. The open and closed field lines are coded with purple and white lines.
The grayscale image denotes the LOS component of the magnetic field at the photosphere.
\label{fig13}}
\end{figure}

\begin{figure}
\epsscale{.50}
\plotone{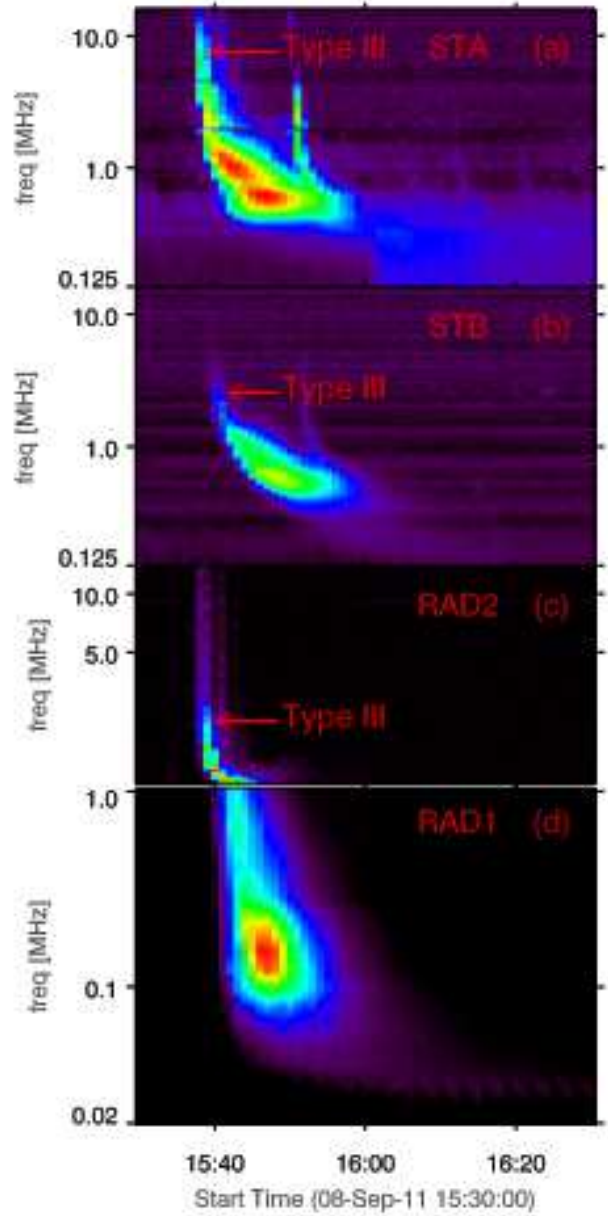}
\caption{Radio dynamic spectra observed by S/WAVES aboard \textit{STA} (panel (a))
and \textit{STB} (panel (b)) and by RAD2 (panel (c)) and RAD1 (panel (d)) aboard
\textit{WIND}/WAVES. The type \Rmnum{3} radio burst that features rapid frequency
drift from high to low values during 15:38$-$16:00 UT is pointed by the red arrows.
\label{fig14}}
\end{figure}

Then, a question is raised: How can the runaway part of the filament successfully escape from the corona and give
rise to a CME? We speculate that open magnetic field lines provide a channel. In order to justify the speculation, we
turn to the large-scale magnetic field calculated by the potential field source surface \citep[PFSS;][]{sch69,sch03}
modelling and the radio dynamic spectra from the S/WAVES
and WAVES instruments. In Figure~\ref{fig13}, we show the magnetic field lines whose footpoints are located in AR 11283 at
12:04 UT before the onset of flare/CME event. The open and closed field lines are represented by the purple and white lines.
It is clear that open field lines do exist in the AR and their configuration accords with the directions of the escaping part of
filament observed by AIA and the CME observed by C2.
The radio dynamic spectra from S/WAVES and WAVES are displayed in panels (a)$-$(b) and (c)$-$(d) of Figure~\ref{fig14},
respectively. There are clear signatures of type \Rmnum{3} radio burst in the spectra. For \textit{STA}, the burst started at
$\sim$15:38:30 UT and ended at $\sim$16:00 UT, during which the frequency drifted rapidly from 16 MHz to $\sim$0.3 MHz.
For \textit{STB} that was $\sim$0.07 AU further than \textit{STA} from the Sun, the burst started slightly later by $\sim$2 minutes
with the frequency drifting from $\sim$4.1 MHz to $\sim$0.3 MHz since the early propagation of the filament was blocked
by the Sun. For WAVES, the burst started at $\sim$15:39:30 UT and ended at $\sim$16:00 UT with the frequency drifting
from 13.8 MHz to $\sim$0.03 MHz. The starting times of the radio burst were consistent with the HXR peak times of the flare.
Since the type \Rmnum{3} radio emissions result from the cyclotron maser instability of the nonthermal electron beams
that are accelerated and ejected into the interplanetary space along open magnetic field lines during the flare \citep{tang13},
the type \Rmnum{3} radio burst observed by \textit{STEREO} and \textit{WIND} provides indirect and supplementary evidence that
open magnetic field lines exist near the flare site.

\section{Discussions} \label{s-disc}

\subsection{How is the energy accumulated?} \label{s-eng}
It is widely accepted that the solar eruptions result from the release of magnetic free energy. For this event, we studied
how the energy is accumulated by investigating the magnetic evolution of the AR using the HMI LOS magnetograms (see the
online movie Animation5.mpg). Figure~\ref{fig15} displays four snapshots of the magnetograms, where the AR is dominated
by negative polarity (N1). A preexisting positive polarity (P1) is located in the northeast direction. From the movie, we found
continuous shearing motion along the highly fragmented and complex PIL between N1 and P1. For example, the small negative
region N2 at the boundary of the sunspot was dragged westward and became elongated (Figure~\ref{fig15}(b)-(d)).
To better illustrate the motion, we derived the transverse velocity field ($v_x$, $v_y$) at the photosphere using the
differential affine velocity estimator (DAVE) method \citep{sch05}. The cadence of the HMI LOS
magnetograms was lowered from 45 s to 180 s. Figure~\ref{fig16} displays six snapshots of the magnetograms overlaid with
the transverse velocity field represented by the white arrows. The velocity field is clearly characterized by the shearing
motions along the PIL. The regions within the green and blue elliptical lines are dominated by
eastward and westward motions at the speeds of $\sim$1.5 km s$^{-1}$. From the online movie (Animation6.mpg), we can see
that the continuous shearing motions were evident before the flare, implying that the magnetic free energy and helicity were
accumulated and stored before the impulsive release.

\subsection{How is the eruption triggered?} \label{s-tri}
\begin{figure}
\epsscale{.90}
\plotone{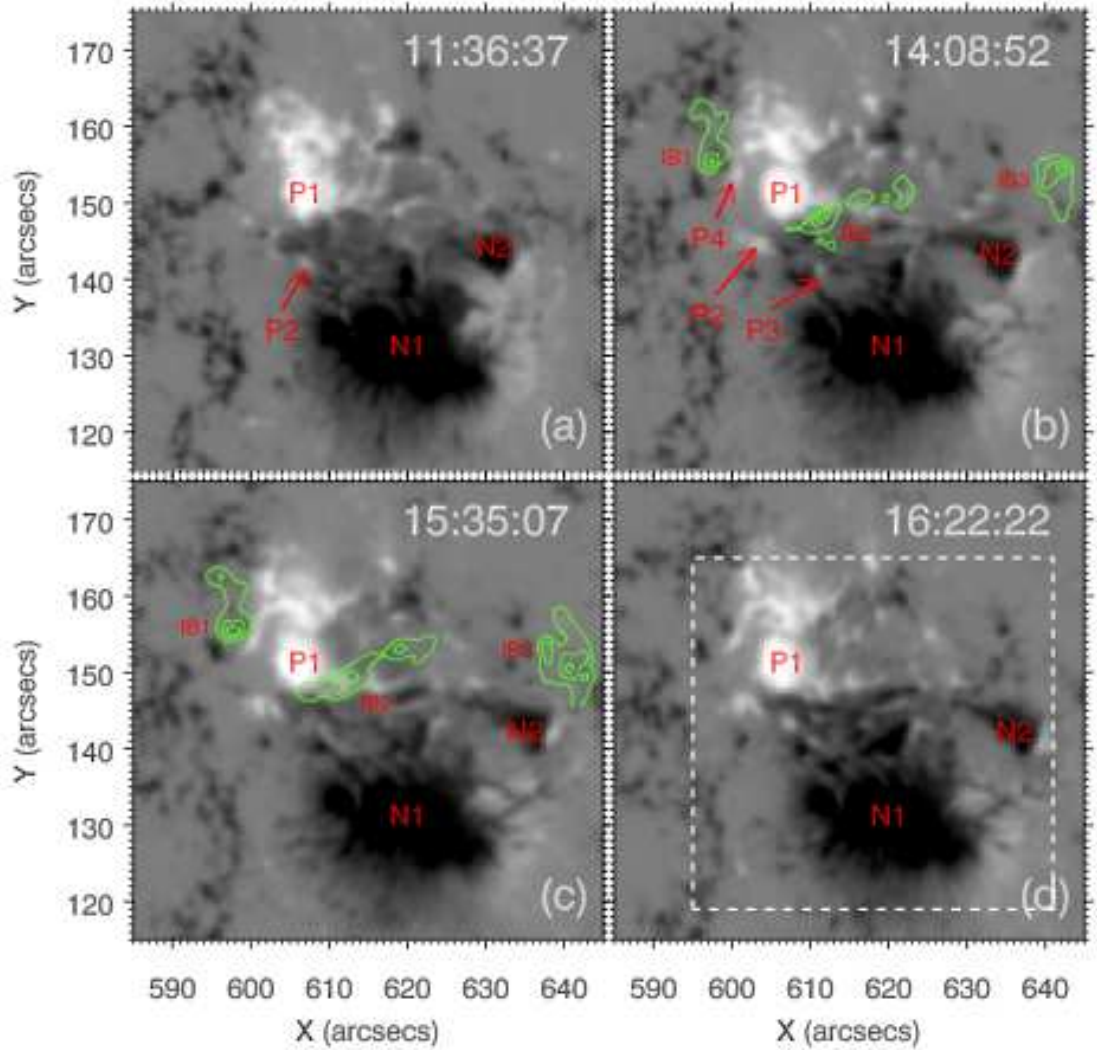}
\caption{(a)$-$(d) Four snapshots of the HMI LOS magnetograms. The AR is dominated by
negative polarities (N1 and N2) that are adjacent to a preexisting (P1) and three emerging (P2,
P3, and P4) positive polarities. The contours of the EUV 304 {\AA} intensities in Figure~\ref{fig3}(b)
and (d) are superposed with green lines in panel (b) and (c), respectively. The total positive and
negative magnetic fluxes within the white dashed box of panel (d) are calculated and their
temporal evolution are plotted in Figure~\ref{fig17}.
\label{fig15}}
(Animations of this figure are available in the online journal.)
\end{figure}

\begin{figure}
\epsscale{.65}
\plotone{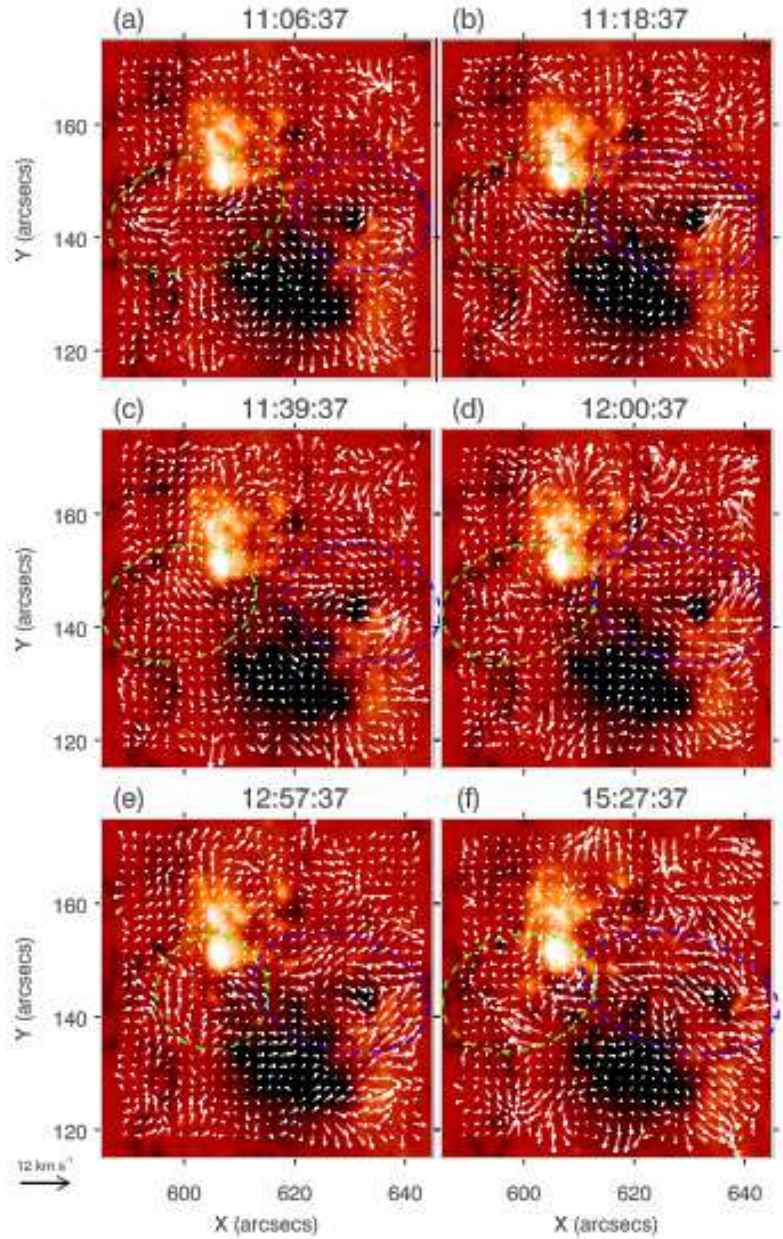}
\caption{(a)$-$(f) Six snapshots of the HMI LOS magnetograms overlaid with the transverse
velocity field represented by the white arrows. The regions within the green (blue) elliptical lines
are dominated by eastward (westward) shearing motions.
\label{fig16}}
(Animations of this figure are available in the online journal.)
\end{figure}

\begin{figure}
\epsscale{.90}
\plotone{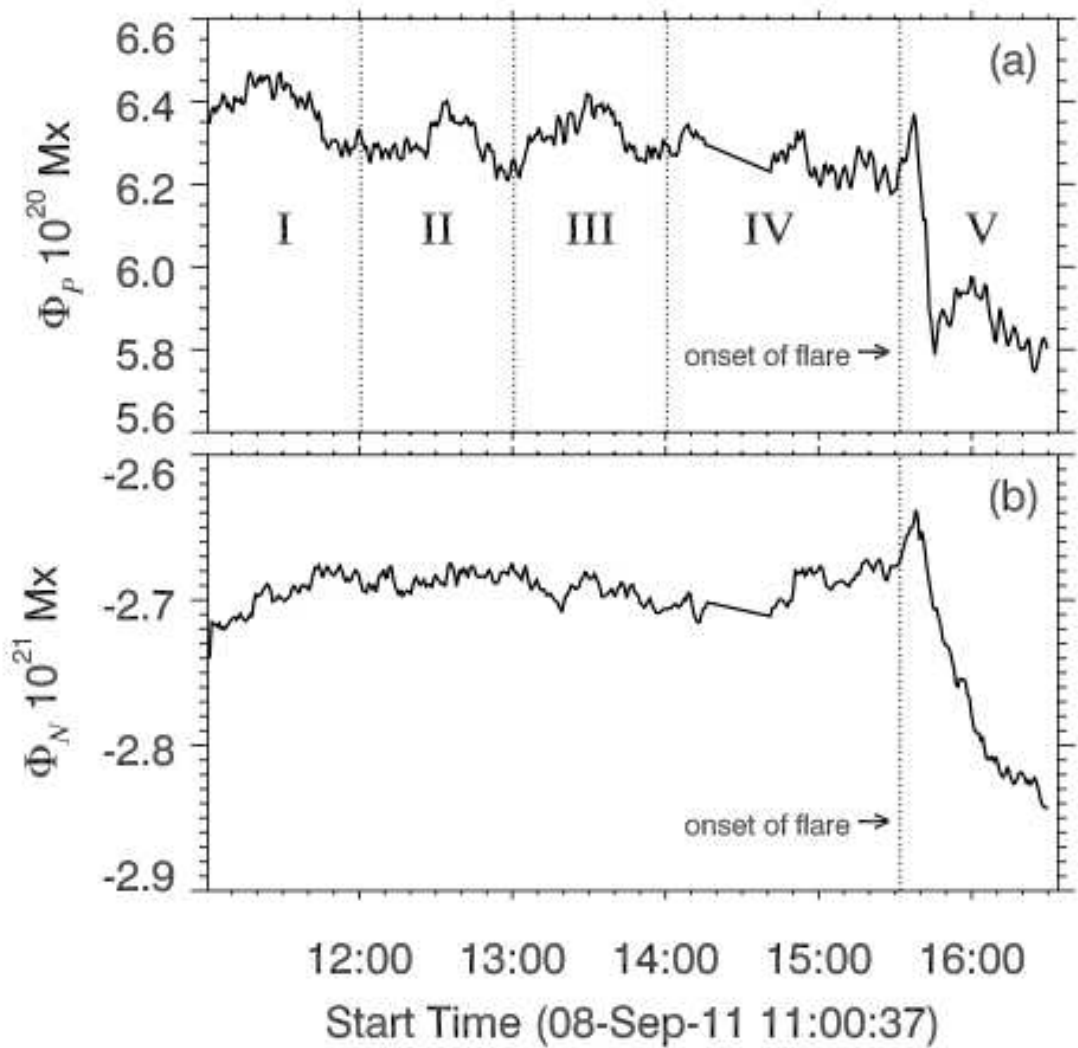}
\caption{(a)$-$(b) Temporal evolutions of the total positive ($\Phi_{P}$) and negative
($\Phi_{N}$) magnetic fluxes within the white dashed box of Figure~\ref{fig15}(d). The
evolution of $\Phi_{P}$ is divided into five phases (I$-$V) separated
by dotted lines. The arrows in both panels point to the starting time of the flare.
\label{fig17}}
\end{figure}

Once the free energy of the AR is accumulated to a critical value, chances are that the filament constrained by the overlying
magnetic field lines undergoes an eruption. Several types of triggering mechanism have been proposed. One type of
processes where magnetic reconnection is involved include the flux emergence model \citep{chen00}, catastrophic model
\citep{lin00}, tether-cutting model \citep{moo01,chen14}, and breakout model \citep{ant99}, to name a few. Another type is the
ideal magnetohydrodynamic (MHD) processes as a result of KI \citep{kli04} and/or TI \citep{kli06}. From Figure~\ref{fig15} and
the movie (Animation5.mpg), we can see that before the flare there was continuous magnetic flux emergence (P2, P3, and P4)
and subsequent magnetic cancellation along the fragmented PIL. We extracted a large region within the white dashed box of
Figure~\ref{fig15}(d) and calculated the total positive ($\Phi_{P}$) and negative ($\Phi_{N}$) magnetic fluxes within the box.
In Figure~\ref{fig17}, the temporal evolutions of the fluxes during 11:00$-$16:30 UT are plotted, with the evolution of $\Phi_{P}$
divided into five phases (I$-$V) separated by the dotted lines. The first four phases before the onset of flare at 15:32 UT are
characterized by quasi-periodic and small-amplitude magnetic flux emergence and cancellation, implying that the large-scale
magnetic field was undergoing rearrangement before the flare.
The intensity contours of the 304 {\AA} images in Figure~\ref{fig3}(b) and (d) are overlaid on the magnetograms in
Figure~\ref{fig15}(b) and (c), respectively. It is clear that the initial brightenings IB1 and IB2 are very close to the small positive
polarities P4 and P3. There is no significant magnetic flux emergence around IB3.
In the emerging-flux-induced-eruption model \citep{chen00}, when reconnection-favorable magnetic bipole emerges from
beneath the photosphere into the filament channel, it reconnects with the preexisting
magnetic field lines that compress the inverse-polarity MFR. The small-scale magnetic reconnection and flux cancellation
serve as the precursor for the upcoming filament eruption and flare. During the flare when magnetic reconnection occurred
between 15:32 UT and 16:10 UT, both the positive and negative magnetic field experienced impulsive and irreversible changes.

Despite that the flux emergences are plausible to interpret the triggering mechanism, there is another possibility.
In the tether-cutting model \citep{moo01}, a pair of $J$-shape sheared arcades that comprise a sigmoid reconnect when
the two elbows come into contact, forming a short loop and a long MFR. Whether the MFR experiences a failed or ejective
eruption depends on the strength of compression from the large-scale background field. The initial brightenings (IB1, IB2, and IB3)
around the sigmoidal filament might be the precursor brightenings as a result of internal tether-cutting reconnection due
to the continuous shearing motion along the PIL. After onset, the whole flux system erupted and produced the M-class flare.
Considering that the magnetic configuration could not be modelled during the flare, we are not sure whether a coherent MFR
was formed after the initiation \citep{chen14}. Compared to the flux emergences, the internal tether-cutting seems more 
believable to interpret how the filament eruption was triggered for the following reasons. Firstly, the filament was supported by 
sheared arcade. Secondly, there were continuous shearing motions along the PIL, and the directions were favorable for the 
tether-cutting reconnection. Finally, the initial brightenings (IB1, IB2, and IB3) around the filament in Figure~\ref{fig3} fairly match 
the internal tether-cutting reconnection with the presence of multiple bright patches of flare emission in the chromosphere at the 
feet of reconnected field lines, while there was no flux emergence around IB3.
NLFFF modelling shows that the twist number ($\sim$1) of the sheared arcades supporting the filament is less than the
threshold value ($\sim$1.5), implying that the filament eruption may not be triggered by ideal KI. The photospheric magnetic
field of the AR features a bipole (P1 and N1) and a couple of mini-polarities (e.g., P2, P3, P4, and N2). Therefore, the filament
eruption could not be explained by the breakout model that requires quadrupolar magnetic field, although null-point magnetic
reconnection took place above the filament during the eruption.

After the onset of eruption, the filament split into two parts as described in Section~\ref{s-result}.
How the filament split is still unclear. In the previous literatures, magnetic reconnection is involved in the split in most cases
\citep{gil01,gib06a,liu08b}. In this study, the split occurred during the impulsive phase of the flare at the eastern leg that was closer
to the flare site than the western one, implying that the split was associated with the release of magnetic energy. The subsequent
rotation or unwinding motion implies the release of magnetic helicity stored in the filament before the flare, presumably due to the
shearing motion in the photosphere. Nevertheless, it is still elusive whether the filament existed as a whole or was composed of two
interwinding parts before splitting. The way of splitting seems difficult to be explained by any of the previous models and requires
in-depth investigations.

Though the runaway part escaped out of the corona, the major part failed. It returned to the solar surface after reaching
the apex. Such kind of failed eruptions have been frequently observed and explained by the strapping effect of the overlying
arcade \citep{ji03,guo10a,song14,jos14} or asymmetry of the background magnetic fields with respect to the location of the filament
\citep{liu09}. In order to figure out the cause of failed
eruption of the major part, we turn to the large-scale magnetic configurations displayed in the bottom panels of Figure~\ref{fig2}.
It is revealed that the overlying magnetic arcades above AR 11283 are asymmetric to a great extent, i.e., the magnetic field to the
west of AR is much stronger than that to the east, which is similar to the case of \citet{liu09}. According to the analysis of \citet{liu09},
the confinements of the large-scale arcade acted on the filament are strong enough to prevent it from escaping. We also performed 
magnetic potential-field extrapolation using the same boundary and derived the distributions of $|\mathbf{B}|$ above the PIL. It is 
found that the maximum height of the major part considerably exceeds the critical height ($\sim$80$\arcsec$) of TI where the decay 
index ($-d\ln|\mathbf{B}|/d\ln z$) of the background potential field reaches $\sim$1.5. The major part would have escaped from the
corona successfully after entering the instability domain if TI had worked. Therefore, the asymmetry with respect to the filament location, 
rather than TI of the overlying arcades, seems reasonable and convincing to interpret why the major part of the filament underwent failed 
eruption. In this study, both successful and failed eruptions occurred in a partially eruptive event, which provides more constraints to the
theoretical models of solar eruptions.

\subsection{How is the coronal loop oscillation triggered?} \label{s-loop}
Since the first discovery of coronal loop oscillations during flares \citep{asch99,nak99}, such kind of oscillations are found to
be ubiquitous and be useful for the diagnostics of coronal magnetic field \citep{guo15}.
Owing to the complex interconnections of the magnetic field lines, blast wave and/or EUV wave induced by filament eruption
may disturb the adjacent coronal loops in the same AR or remote loops in another AR, resulting in transverse kink-mode
oscillations. \citet{nis13} observed decaying and decayless transverse oscillations of a coronal loop on 2012 May 30. 
The loops experience small-amplitude decayless oscillations, which is driven by an external non-resonant harmonic driver 
before and after the flare \citep{mur14}. The flare, as an impulsive driver, triggers large-amplitude decaying loop oscillations. 
In our study, the decayless loop oscillation with moderate amplitude ($\sim$1.6 Mm) occurred during the flare and lasted for 
only two cycles, which makes it quite difficult to precisely measure the decay timescale if it is decaying indeed. The loop may 
cool down and become invisible in 171 {\AA} while oscillating. Considering that the distance between the flare and
OL is $\sim$50 Mm and the time delay between the flare onset and loop oscillation is $\sim$6 minutes, the speed of
propagation of the disturbances from the flare to OL is estimated to be $\sim$140 km s$^{-1}$, which is close to the local
sound speed of the plasmas with temperature of $\sim$0.8 MK. Hence, we suppose that the coronal loop oscillation was
triggered by the external disturbances as a result of the rising and expanding motions of the filament.

\subsection{Significance for space weather prediction} \label{s-swp}
Flares and CMEs play a very important role in the generation of space weather. Accurate prediction
of space weather is of great significance. Successful eruptions have substantially been observed and deeply investigated.
Partial filament eruptions that produce flares and CMEs, however, are rarely detected and poorly explored. For the type
of partial eruptions in this study, i.e., one part undergoes failed eruption and the other part escapes out of the corona, it
would be misleading and confusing to assess and predict the space weather effects based on the information only from
the solar surface, since the escaping part may carry or produce solar energetic particles that have potential geoeffectiveness.
Complete observations are necessary for accurate predictions.

\section{Summary} \label{s-sum}
Using the multiwavelength observations from both spaceborne and ground-based telescopes, we
studied in detail a partial filament eruption event in AR 11283 on 2011 September 8. The main results
are summarized as follows:

\begin{enumerate}
  \item{A magnetic null point was found above the preexisting positive polarity surrounded by
  negative polarities in the AR. A spine passed through the null and intersected with the photosphere
  to the left. Weakly twisted sheared arcade supporting the filament was located under the null point
  whose height increased slightly by $\sim$0.4 Mm after the eruption.}
  \item{The filament rose and expanded, which was probably triggered by the internal tether-cutting
  reconnection or by continuous magnetic flux emergence and cancellation along the highly complex
  and fragmented PIL, the former of which seems more convincing. 
  During its eruption, it triggered the null-point magnetic reconnection
  and the M6.7 flare with a single HXR source at different energy bands. The flare produced
  a quasi-circular ribbon and a V-shape ribbon where the outer spine intersects with the photosphere.}
  \item{During the expansion, the filament split into two parts at the eastern leg that is closer to the flare
  site. The major part of the filament rose at the speeds of 90$-$150 km s$^{-1}$ before reaching the
  maximum apparent height of $\sim$115 Mm. Afterwards, it returned to the solar surface staggeringly
  at the speeds of 20$-$80 km s$^{-1}$. The rising and falling motions of the filament were clearly observed
  in the UV, EUV, and H$\alpha$ wavelengths. The failed eruption of the major part was most probably
  caused by the asymmetry of the overlying magnetic arcades with respect to the filament location.}
  \item{The runaway part, however, separated from and rotated around the major part for $\sim$1 turn
  before escaping outward from the corona at the speeds of 125$-$255 km s$^{-1}$, probably along
  the large-scale open magnetic field lines as evidenced by the PFSS modelling and the type \Rmnum{3}
  radio burst. The ejected part of the filament led to a faint CME. The angular width and apparent speed
  of the CME in the FOV of C2 are 37$^{\circ}$ and 214 km s$^{-1}$. The propagation directions of the
  escaping filament observed by SDO/AIA and \textit{STA}/EUVI are consistent with those of the CME
  observed by LASCO/C2 and \textit{STA}/COR1, respectively.}
  \item{The partial filament eruption also triggered transverse oscillation of the neighbouring
  coronal loops in the same AR. The amplitude and period of the kink-mode oscillation were 1.6 Mm
  and 225 s. We also performed diagnostics of the plasma density and temperature of the
  oscillating loops.}
\end{enumerate}

\acknowledgements The authors thank the referee for valuable suggestions and comments to
improve the quality of this article. We gratefully acknowledge Y. N. Su, P. F. Chen, J. Zhang, B. Kliem,
R. Liu, S. Gibson, H. Gilbert, M. D. Ding, and H. N. Wang for inspiring and constructive discussions.
\textit{SDO} is a mission of NASA\rq{}s Living With a Star Program. AIA and HMI data are courtesy
of the NASA/\textit{SDO} science teams.
\textit{STEREO}/SECCHI data are provided by a consortium of US, UK, Germany, Belgium, and
France. QMZ is supported by Youth Fund of JiangSu BK20141043, by 973 program under grant
2011CB811402, and by NSFC 11303101, 11333009, 11173062, 11473071, and 11221063.
H. Ji is supported by the Strategic Priority Research Program$-$The Emergence of
Cosmological Structures of the Chinese Academy of Sciences, Grant No. XDB09000000.
YG is supported by NSFC 11203014.
Li Feng is supported by the NFSC grant 11473070, 11233008 and by grant BK2012889.
Li Feng also thanks the Youth Innovation Promotion Association, CAS, for the financial support.

\clearpage

\clearpage

\clearpage

\clearpage

\clearpage

\clearpage

\clearpage

\clearpage

\clearpage

\clearpage

\clearpage

\clearpage

\clearpage

\clearpage

\clearpage

\clearpage

\clearpage

\clearpage

\begin{table}
\caption{Description of the observational parameters.}
\label{tbl-1}
\centering
\begin{tabular}{l c c c c}
\hline\hline
Instrument & $\lambda$ & Time & Cadence & Pixel Size \\
           &     ({\AA})   & (UT)  &    (sec)    &  (arcsec)  \\
\hline
  BBSO & 6563 & 15:30$-$17:00 & 60  & 1.0 \\
  AIA & 94$-$1700 & 15:30$-$16:30 & 12 & 0.6 \\
  HMI &  6173        & 11:00$-$17:00 & 45, 720 & 0.5 \\
  EUVI & 195  & 15:30$-$16:30 & 300 & 1.6 \\
  EUVI & 304  & 15:30$-$16:30 & 600 & 1.6 \\
  COR1 & WL & 15:30$-$16:30 & 300 & 15.0 \\
  S/WAVES & 2.5$-$16025 kHz & 15:30$-$16:30 & 60 & $-$ \\
  WAVES & 20$-$13825 kHz & 15:30$-$16:30 & 60 & $-$ \\
  LASCO/C2 & WL & 15:30$-$18:00 & 720 & 11.4 \\
  \textit{GOES}   & 0.5$-$4.0  & 15:30$-$16:30 & 3  & $-$ \\
  \textit{GOES}   & 1.0$-$8.0  &  15:30$-$16:30 & 3  & $-$  \\
  \textit{RHESSI} & 3$-$100 keV & 15:30$-$16:10 & 4, 10  & 4  \\
\hline
\end{tabular}
\end{table}

\clearpage


\begin{thebibliography}{}
\bibitem[Amari et al.(2014)]{ama14} Amari, T., Canou, A., \& Aly, J.-J.\ 2014, \nat, 514, 465
\bibitem[Antiochos et al.(1999)]{ant99} Antiochos, S.~K., DeVore, C.~R., \& Klimchuk, J.~A.\ 1999, \apj, 510, 485
\bibitem[Aschwanden et al.(1999)]{asch99} Aschwanden, M.~J., Fletcher, L., Schrijver, C.~J.,
\& Alexander, D.\ 1999, \apj, 520, 880
\bibitem[Aschwanden(2011)]{asch11} Aschwanden, M.~J.\ 2011, Living Reviews in Solar Physics, 8, 5
\bibitem[Bi et al.(2014)]{bi14} Bi, Y., Jiang, Y., Yang, J., et al.\ 2014, \apj, 790, 100
\bibitem[Bougeret et al.(1995)]{bou95} Bougeret, J.-L.,
Kaiser, M.~L., Kellogg, P.~J., et al.\ 1995, \ssr, 71, 231
\bibitem[Bougeret et al.(2008)]{bou08} Bougeret, J.~L., Goetz, K., Kaiser, M.~L., et al.\ 2008, \ssr, 136, 487
\bibitem[Brueckner et al.(1995)]{bru95} Brueckner, G.~E., Howard, R.~A., Koomen, M.~J., et al.\ 1995,
\solphys, 162, 357
\bibitem[Chen(2011)]{chen11} Chen, P.~F.\ 2011, Living Reviews in Solar Physics, 8, 1
\bibitem[Chen \& Shibata(2000)]{chen00} Chen, P.~F., \& Shibata, K.\ 2000, \apj, 545, 524
\bibitem[Chen et al.(2008)]{chen08} Chen, P.~F., Innes, D.~E., \& Solanki, S.~K.\ 2008, \aap, 484, 487
\bibitem[Chen et al.(2014)]{chen14} Chen, H., Zhang, J., Cheng, X., et al.\ 2014, \apjl, 797, LL15
\bibitem[Cheng et al.(2012)]{cx12} Cheng, X., Zhang, J., Saar, S.~H., \& Ding, M.~D.\ 2012, \apj, 761, 62
\bibitem[Cheng et al.(2014a)]{cx14a} Cheng, X., Ding, M.~D., Zhang, J., et al.\ 2014a, \apjl, 789, L35
\bibitem[Cheng et al.(2014b)]{cx14b} Cheng, X., Ding, M.~D., Zhang, J., et al.\ 2014b, \apj, 789, 93
\bibitem[Dai et al.(2013)]{dai13} Dai, Y., Ding, M.~D., \& Guo, Y.\ 2013, \apjl, 773, L21
\bibitem[Fan(2005)]{fan05} Fan, Y.\ 2005, \apj, 630, 543
\bibitem[Feng et al.(2013)]{feng13} Feng, L., Wiegelmann, T., Su, Y., et al.\ 2013, \apj, 765, 37
\bibitem[Gibson \& Fan(2006a)]{gib06a} Gibson, S.~E., \& Fan, Y.\ 2006a, \apjl, 637, L65
\bibitem[Gibson \& Fan(2006b)]{gib06b} Gibson, S.~E., \& Fan, Y.\ 2006b,
Journal of Geophysical Research (Space Physics), 111, 12103
\bibitem[Gilbert et al.(2000)]{gil00} Gilbert, H.~R., Holzer, T.~E., Burkepile, J.~T., \& Hundhausen, A.~J.\ 2000, \apj, 537, 503
\bibitem[Gilbert et al.(2001)]{gil01} Gilbert, H.~R., Holzer, T.~E., \& Burkepile, J.~T.\ 2001, \apj, 549, 1221
\bibitem[Gilbert et al.(2007)]{gil07} Gilbert, H.~R., Alexander, D., \& Liu, R.\ 2007, \solphys, 245, 287
\bibitem[Guo et al.(2010a)]{guo10a} Guo, Y., Ding, M.~D., Schmieder, B., et al.\ 2010a, \apjl, 725, L38
\bibitem[Guo et al. (2010b)]{guo10b} Guo, Y., Schmieder, B., D{\'e}moulin, P., et al.\ 2010b, \apj, 714, 343
\bibitem[Guo et al.(2013)]{guo13} Guo, Y., D{\'e}moulin, P., Schmieder, B., et al.\ 2013, \aap, 555, A19
\bibitem[Guo et al.(2015)]{guo15} Guo, Y., Erd{\'e}lyi, R., Srivastava, A.~K., et al.\ 2015, \apj, 799, 151
\bibitem[Hood \& Priest(1981)]{hood81} Hood, A.~W., \& Priest, E.~R.\ 1981,
Geophysical and Astrophysical Fluid Dynamics, 17, 297
\bibitem[Howard et al.(2008)]{how08} Howard, R.~A., Moses, J.~D., Vourlidas, A., et al.\ 2008, \ssr, 136, 67
\bibitem[Ji et al.(2003)]{ji03} Ji, H., Wang, H., Schmahl, E.~J., Moon, Y.-J., \& Jiang, Y.\ 2003, \apjl, 595, L135
\bibitem[Jiang et al.(2013)]{jiang13} Jiang, C., Feng, X., Wu, S.~T., \& Hu, Q.\ 2013, \apjl, 771, L30
\bibitem[Jiang et al.(2014)]{jiang14} Jiang, C., Wu, S.~T., Feng, X., \& Hu, Q.\ 2014, \apj, 780, 55
\bibitem[Joshi et al.(2013)]{jos13} Joshi, N.~C., Srivastava, A.~K., Filippov, B., et al.\ 2013, \apj, 771, 65
\bibitem[Joshi et al.(2014)]{jos14} Joshi, N.~C., Srivastava, A.~K., Filippov, B., et al.\ 2014, \apj, 787, 11
\bibitem[Kaiser(2005)]{kai05} Kaiser, M.~L.\ 2005, Advances in Space Research, 36, 1483
\bibitem[Keppens \& Xia(2014)]{rony14} Keppens, R., \& Xia, C.\ 2014, \apj, 789, 22
\bibitem[Kliem et al.(2004)]{kli04} Kliem, B., Titov, V.~S., T{\"o}r{\"o}k, T.\ 2004, \aap, 413, L23
\bibitem[Kliem \& T{\"o}r{\"o}k(2006)]{kli06} Kliem, B., T{\"o}r{\"o}k, T.\ 2006, Physical Review Letters, 96, 255002
\bibitem[Kumar et al.(2010)]{kur10} Kumar, P., Srivastava, A.~K., Filippov, B., \& Uddin, W.\ 2010, \solphys, 266, 39
\bibitem[Kumar et al.(2011)]{kur11} Kumar, P., Srivastava, A.~K., Filippov, B., Erd{\'e}lyi, R.,
\& Uddin, W.\ 2011, \solphys, 272, 301
\bibitem[Kumar et al.(2012)]{kur12} Kumar, P., Cho, K.-S., Bong, S.-C., Park, S.-H., \& Kim, Y.~H.\ 2012, \apj, 746, 67
\bibitem[Lemen et al.(2012)]{lem12} Lemen, J.~R., Title, A.~M., Akin, D.~J., et al.\ 2012, \solphys, 275, 17
\bibitem[Li \& Zhang(2012)]{li12} Li, T., \& Zhang, J.\ 2012, \apjl, 760, L10
\bibitem[Li et al.(2014)]{li14} Li, Y., Ding, M.~D., Guo, Y., \& Dai, Y.\ 2014, \apj, 793, 85
\bibitem[Lin et al.(2002)]{lin02} Lin, R.~P., Dennis, B.~R., Hurford, G.~J., et al.\ 2002, \solphys, 210, 3
\bibitem[Lin \& Forbes(2000)]{lin00} Lin, J., \& Forbes, T.~G.\ 2000, \jgr, 105, 2375
\bibitem[Liu et al.(2007)]{liu07} Liu, R., Alexander, D., \& Gilbert, H.~R.\ 2007, \apj, 661, 1260
\bibitem[Liu(2008)]{liu08a} Liu, Y.\ 2008, \apjl, 679, L151
\bibitem[Liu et al.(2008)]{liu08b} Liu, R., Gilbert, H.~R., Alexander, D., \& Su, Y.\ 2008, \apj, 680, 1508
\bibitem[Liu et al.(2009)]{liu09} Liu, Y., Su, J., Xu, Z., et al.\ 2009, \apjl, 696, L70
\bibitem[Liu et al.(2012)]{liu12} Liu, R., Kliem, B., T{\"o}r{\"o}k, T., et al.\ 2012, \apj, 756, 59
\bibitem[Liu et al.(2014)]{liu14} Liu, C., Deng, N., Lee, J., et al.\ 2014, \apj, 795, 128
\bibitem[Mackay et al.(2010)]{mac10} Mackay, D.~H., Karpen, J.~T., Ballester, J.~L., Schmieder, B.,
\& Aulanier, G.\ 2010, \ssr, 151, 333
\bibitem[Mandrini et al.(2014)]{man14} Mandrini, C.~H., Schmieder, B., D{\'e}moulin, P., Guo, Y.,
\& Cristiani, G.~D.\ 2014, \solphys, 289, 2041
\bibitem[Moore et al.(2001)]{moo01} Moore, R.~L., Sterling, A.~C., Hudson, H.~S., \& Lemen, J.~R.\ 2001, \apj, 552, 833
\bibitem[Murawski et al.(2014)]{mur14} Murawski, K., Solov'ev, A., Kraskiewicz, J., \& Srivastava, A.~K.\ 2014, arXiv:1411.7465
\bibitem[Nakariakov et al.(1999)]{nak99} Nakariakov, V.~M., Ofman, L., Deluca, E.~E., Roberts, B.,
\& Davila, J.~M.\ 1999, Science, 285, 862
\bibitem[Nakariakov \& Ofman(2001)]{nak01} Nakariakov, V.~M., \& Ofman, L.\ 2001, \aap, 372, L53
\bibitem[Ning \& Cao(2010)]{ning10} Ning, Z., \& Cao, W.\ 2010, \solphys, 264, 329
\bibitem[Nistic{\`o} et al.(2013)]{nis13} Nistic{\`o}, G., Nakariakov, V.~M., \& Verwichte, E.\ 2013, \aap, 552, AA57
\bibitem[Ruan et al.(2014)]{ruan14} Ruan, G., Chen, Y., Wang, S., et al.\ 2014, \apj, 784, 165
\bibitem[Schatten et al.(1969)]{sch69} Schatten, K.~H., Wilcox, J.~M., \& Ness, N.~F.\ 1969, \solphys, 6, 442
\bibitem[Scherrer et al.(2012)]{sch12} Scherrer, P.~H., Schou, J., Bush, R.~I., et al.\ 2012, \solphys, 275, 207
\bibitem[Schrijver \& De Rosa(2003)]{sch03} Schrijver, C.~J., \& De Rosa, M.~L.\ 2003, \solphys, 212, 165
\bibitem[Schuck(2005)]{sch05} Schuck, P.~W.\ 2005, \apjl, 632, L53
\bibitem[Shen et al.(2012)]{shen12} Shen, Y., Liu, Y., \& Su, J.\ 2012, \apj, 750, 12
\bibitem[Shen et al.(2014)]{shen14} Shen, Y., Liu, Y.~D., Chen, P.~F., \& Ichimoto, K.\ 2014, \apj, 795, 130
\bibitem[Song et al.(2014)]{song14} Song, H.~Q., Zhang, J., Cheng, X., et al.\ 2014, \apj, 784, 48
\bibitem[Srivastava et al.(2010)]{sri10} Srivastava, A.~K., Zaqarashvili, T.~V., Kumar, P.,
\& Khodachenko, M.~L.\ 2010, \apj, 715, 292
\bibitem[Su \& van Ballegooijen(2012)]{su12} Su, Y., \& van Ballegooijen, A.\ 2012, \apj, 757, 168
\bibitem[Sun et al.(2012a)]{sun12a} Sun, X., Hoeksema, J.~T., Liu, Y., et al.\ 2012a, \apj, 748, 77
\bibitem[Sun et al.(2012b)]{sun12b} Sun, X., Hoeksema, J.~T., Liu, Y., Chen, Q., \& Hayashi, K.\ 2012b, \apj, 757, 149
\bibitem[Tang et al.(2013)]{tang13} Tang, J.~F., Wu, D.~J., \& Tan, C.~M.\ 2013, \apj, 779, 83
\bibitem[Terradas et al.(2015)]{ter15} Terradas, J., Soler, R., Luna, M., Oliver, R., \& Ballester, J.~L.\ 2015, \apj, 799, 94
\bibitem[T{\"o}r{\"o}k et al.(2004)]{tor04} T{\"o}r{\"o}k, T., Kliem, B., \& Titov, V.~S.\ 2004, \aap, 413, L27
\bibitem[T{\"o}r{\"o}k \& Kliem(2005)]{tor05} T{\"o}r{\"o}k, T., \& Kliem, B.\ 2005, \apjl, 630, L97
\bibitem[T{\"o}r{\"o}k et al.(2010)]{tor10} T{\"o}r{\"o}k, T., Berger, M.~A., \& Kliem, B.\ 2010, \aap, 516, A49
\bibitem[Tripathi et al.(2006)]{tri06} Tripathi, D., Solanki, S.~K., Schwenn, R., et al.\ 2006, \aap, 449, 369
\bibitem[Tripathi et al.(2007)]{tri07} Tripathi, D., Solanki, S.~K., Mason, H.~E., \& Webb, D.~F.\ 2007, \aap, 472, 633
\bibitem[Tripathi et al.(2009)]{tri09} Tripathi, D., Gibson, S.~E., Qiu, J., et al.\ 2009, \aap, 498, 295
\bibitem[Tripathi et al.(2013)]{tri13} Tripathi, D., Reeves, K.~K., Gibson, S.~E., Srivastava, A.,
\& Joshi, N.~C.\ 2013, \apj, 778, 142
\bibitem[Wheatland et al.(2000)]{wht00} Wheatland, M.~S., Sturrock, P.~A., \& Roumeliotis, G.\ 2000, \apj, 540, 1150
\bibitem[White \& Verwichte(2012)]{whi12} White, R.~S., \& Verwichte, E.\ 2012, \aap, 537, AA49
\bibitem[Wiegelmann(2004)]{wig04} Wiegelmann, T.\ 2004, \solphys, 219, 87
\bibitem[Xia et al.(2012)]{xia12} Xia, C., Chen, P.~F., \& Keppens, R.\ 2012, \apjl, 748, L26
\bibitem[Xia et al.(2011)]{xia11} Xia, C., Chen, P.~F., Keppens, R., \& van Marle, A.~J.\ 2011, \apj, 737, 27
\bibitem[Xia et al.(2014a)]{xia14a} Xia, C., Keppens, R., Antolin, P., \& Porth, O.\ 2014a, \apjl, 792, L38
\bibitem[Xia et al.(2014b)]{xia14b} Xia, C., Keppens, R., \& Guo, Y.\ 2014b, \apj, 780, 130
\bibitem[Zhang et al.(2012a)]{zhang12a} Zhang, J., Cheng, X., \& Ding, M.-D.\ 2012a, Nature Communications, 3, 747
\bibitem[Zhang et al.(2012b)]{zhang12b} Zhang, Q.~M., Chen, P.~F., Xia, C., \& Keppens, R.\ 2012b, \aap, 542, A52
\bibitem[Zhang et al.(2012c)]{zhang12c} Zhang, Q.~M., Chen, P.~F., Guo, Y., Fang, C., \& Ding, M.~D.\ 2012c, \apj, 746, 19
\bibitem[Zhang et al.(2013)]{zhang13} Zhang, Q.~M., Chen, P.~F., Xia, C., Keppens, R., \& Ji, H.~S.\ 2013, \aap, 554, A124
\bibitem[Zhou et al.(2014)]{zhou14} Zhou, Y.-H., Chen, P.-F., Zhang, Q.-M.,
\& Fang, C.\ 2014, Research in Astronomy and Astrophysics, 14, 581
\end{thebibliography}
\end{document}